## DFT-informed Design of Radiation-Resistant Dilute Ternary Cu Alloys

Vaibhav Vasudevan[1]*, Thomas Schuler[2], Pascal Bellon[1], Robert Averback[1]
[1] Materials Research Laboratory, University of Illinois Urbana-Champaign 104 South Goodwin Ave. MC-230 Urbana, IL 61801
[2] Université Paris-Saclay, CEA, Service de recherche en Corrosion et Comportement des Matériaux, SRMP, 91191 Gif-sur-Yvette, France

*vv22@illinois.edu

## Abstract:

This research establishes a systematic, high-throughput computational framework for designing radiation-resistant, dilute ternary copper-based alloys by addition of solutes that bind to vacancies and reduce their mobility, thus promoting interstitial-vacancy recombination. The first challenge in developing alloys by this method is mitigating the vacancy-mediated solute drag effect, since density functional theory (DFT) calculations show that solutes that bind strongly to vacancies are also rapidly dragged to point-defect sinks, and thus removed from the matrix. To overcome this issue, two types of solutes are added to the Cu matrix: A first solute with a strong vacancy binding energy (*B*-type species) and another solute that binds to 'B' and is a slow diffuser in Cu (*C*-type species). Using DFT, 21 synergistic solute pairs are screened, with 'B'=Zr, Ge, Sn and 'C'=Fe, Co, Mo, Ni, Nb, W, Cr. Two promising alloys, Cu(Zr,Co) and Cu(Zr,Fe) are then investigated in detail in the dilute regime. Diffusion and solute drag in these alloys are modeled using the kinetic cluster expansion approach (KineCluE) under irradiation conditions. It is shown that strong Zr-'C' thermodynamic binding, especially between Zr and Co, significantly reduces the mobility of Zr solute and suppresses the vacancy-mediated solute drag. Using an analytical framework for the standard five-jump frequency model for diffusion in binary alloys, it is found that vacancy-Zr-Co triplets disrupt the kinetic circuits that promote solute drag in the binary alloy by raising the dissociation barrier for the vacancy from the solute.

## I. INTRODUCTION:

Advanced nuclear reactors require the development of novel alloys capable of withstanding extreme irradiation environments for extended durations. A primary cause of material degradation under irradiation is the prolific generation of point defects, e.g., vacancies (V) and self-interstitials, which migrate over long distances [1,2]. This migration is often coupled with solute fluxes, leading to radiation-induced segregation and precipitation (RIS and RIP), which often compromises the alloy's mechanical properties and corrosion resistance [1,3–7].

To counteract such radiation damage in alloys, two main strategies have emerged. A first involves engineering materials with a high density of point defect sinks, such as nanograined materials or alloys containing nano-precipitates (i.e., oxide dispersion strengthened alloys) [8–10]. A key drawback, however, is the potential for these nanostructures to coarsen and lose their effectiveness over time, especially under high-temperature irradiation.

A second common strategy, which is the focus of this work, aims to promote point-defect recombination by limiting their mobility. This is typically achieved through strategic solute additions that bind strongly to point defects, effectively trapping them and increasing the probability of recombination before they can migrate to sinks [11,12]. Experimental results have shown that small additions (~2at%) of oversized solutes such as Ge and Sn can suppress the growth of dislocation loops in copper at higher temperatures under electron irradiation [13]. Other promising results have been reported in stainless steels by incorporating oversized solutes such as Zr and Ti to suppress RIS at low doses of electron and proton irradiation, preventing the detrimental Cr-depletion at grain boundaries (GBs) [14–16]. As the irradiation dose increases to ≈10 dpa, however, RIS returns and GB Cr-depletion reaches levels similar to those measured in unmodified alloys [13,14]. A rationalization of this behavior is that solutes that binds strongly to vacancies ($E_{binding}^{B-V}$>3kT) are also fast diffusers, and thus most susceptible to being dragged by the vacancies toward sinks [17–19].

A parametric kinetic Monte Carlo study by S. Jana et al. [20] introduced a novel approach to overcome this shortcoming where two synergistic solutes are employed in a model alloy system: a *B*-type solute with high vacancy-binding energy, paired with a *C*-type solute that binds strongly to the *B*-type solute and exhibits slow diffusion in the matrix. In the following, for simplicity, these two types of species will be here referred to as *B* and C solutes, since no confusion could exist here with boron and carbon species. The underlying hypothesis is that strong *B-C* interactions lead to the formation of stable, low-mobility solute clusters capable of trapping vacancies effectively in V-*B-C* clusters, countering the solute-drag issues prevalent in binary alloys. While potential ternary alloys in Cu, Ni, and Al are identified in ref. [20] using existing binary thermodynamic data, atomistic calculations are needed to (1) quantify the interactions between *B* and *C* type solutes and vacancies for specific ternary systems [21,22] and (2) assess the impact of these interactions on diffusion of solute and vacancies, and on solute drag. The first objective can be best approached through first-principles density functional theory (DFT) calculations. DFT calculations and associated databases have been instrumental for studying diffusion in dilute binary alloys, though these efforts have largely focused on two-body interactions (e.g., vacancy-solute or interstitial-solute migration) [23–27]. While solute-solute interactions are beginning to be studied using DFT [4,21,25,28], their effect on associated kinetic phenomena such as solute drag and solute diffusivity have not been studied in detail yet owing to their inherent complexity.

This paper provides a comprehensive, first-principles-based assessment of how synergistic solutes in a dilute ternary copper alloy can mitigate vacancy-mediated solute drag. A high-throughput methodology is employed, starting with identifying promising *B* (strong vacancy-binding) and *C* (slow-diffusing) solutes from existing binary diffusion data shown in Figure 1 [27]. Using DFT calculations, candidate *B*-*C* solute pairs are screened for strong thermodynamic binding that would enhance the formation of stable solute-vacancy complexes and thus promote vacancy–interstitial recombination under irradiation. The KineCluE code [4] and further DFT migration energy calculations are then employed to model and analyze the full set of Onsager transport coefficients in selected ternary alloys, here Cu(Zr,Fe) and Cu(Zr,Co).

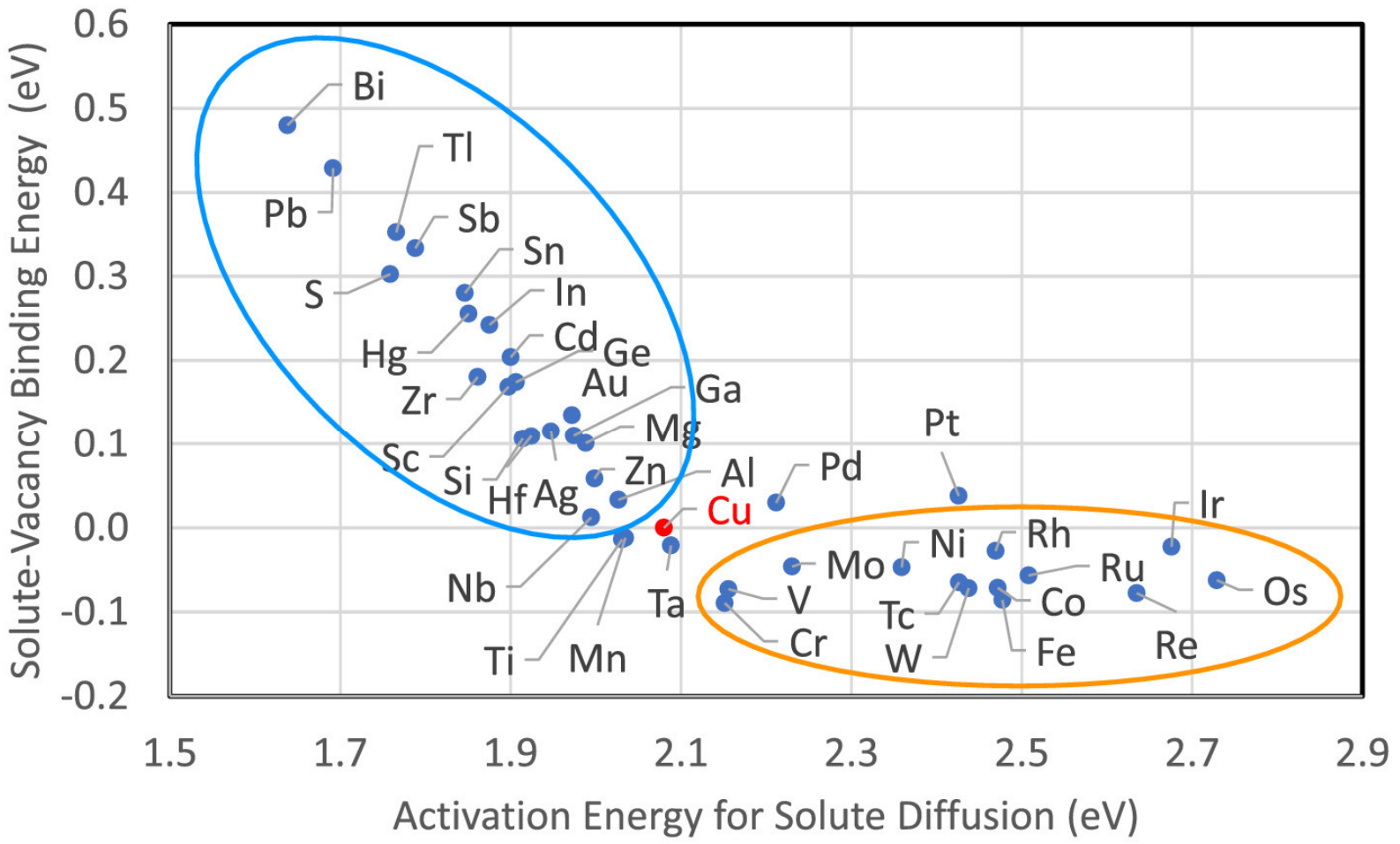

Figure 1: Vacancy solute binding vs. activation energy for solute diffusion. Strong vacancy binding solutes (*B*-type) marked by the blue circle and slow diffusing solutes (*C*-type) marked by the orange circle. Data is adapted from Ref. [27]. Plot is from Ref. [20].

The manuscript is structured as follows: Section 2 details the computational methods, encompassing DFT procedures for calculating binding energies and migration barriers, as well as the KineCluE methodology for calculating transport coefficients. Results from the solute screening, transport coefficient calculations, and flux coupling calculations in dilute ternary Cu alloys are reported in Section 3. The results are then analyzed and discussed in Section 4.

## II. METHODS:

### A. Density Functional Theory (DFT) Calculations:

Density functional theory (DFT) calculations are performed using open-source Quantum Espresso v7.3.1 [29,30]. The exchange and correlation energy was computed with the Perdew-Burke-Ernzerhof (PBE) functional [31] with the generalized gradient approximation (GGA) [32]. Collinear spin-polarized calculations are used for alloys involving magnetic solutes (Fe, Cr¸ Ni or Co). PS-Library pseudopotentials [33] are utilized for all elements using the Plane Augmented Wave (PAW) versions [34]. Calculations are performed using a 3x3x3 FCC supercell containing up to 108 atoms with a 4x4x4 k-point mesh. First-order Methfessel-Paxton smearing of 0.027 eV ensures accurate evaluation of forces for ionic relaxation, see Supplemental Materials [35]. The plane-wave energy cutoff was set to 1020 eV and charge density cutoff was set to 6123 eV, ensuring total energy convergence below 1 meV/atom. The migration energies are calculated using a climbing image nudged elastic band (CI-NEB) method with one intermediate image. From previous literature [17,27,36], one climbing image is found to be suitable for finding the saddle points in FCC crystal structures due to well-defined crystallographic sites for migrating atoms. All images are relaxed until the forces on each ion are converged below 5 meV/Å. The Aneto code was utilized to correct for periodic interactions between solute-defect clusters for both binding energy and migration energy calculations [37]. With a 3x3x3 supercell, 5NN binding energies could become unphysical due to periodic boundary conditions (i.e., 5NN interactions from the first defect are also 5NN through the periodic boundary). However, these interactions through the periodic boundary are small even at 5NN distances due to magnitude of the ANETO corrections, which corrects for the periodic boundary elastic interactions. The 5NN interactions within a 108-atom supercell can be thought of as a <310> elastic dipole interaction which can be corrected for using ANETO. The ANETO correction for 5NN Zr Fe pairs is 4 meV which is small in magnitude compared to the 1-5NN binding energies between the Zr-Fe pairs. The Atomic Simulation Environment (ASE) Python package is used for setting up and managing the calculations [38]. Binding energies between vacancies and/or solute in the copper FCC matrix are calculated using Equation (1) with the convention that positive binding energies are attractive.

$$E_b(A_1, A_2, \dots A_n) = \sum_{i=1,2\dots n} E(A_i) - [(n-1) * E(ref) + E(A_1 + A_2 + \cdots + A_n)] \quad (1)$$

where, $E(A_i)$ is the energy of the supercell containing an isolated defect or solute $A_i$, $E(ref)$ is the reference supercell energy, i.e., containing no defects or solutes, and $E(A_1 + A_2 + \cdots + A_n)$ is the energy of a supercell containing a defect-solute cluster. B-type solutes investigated are Zr, Ge, and Sn. The *C*-type solutes investigated are Fe, Co, Mo, Ni, Nb, Cr, and W. To determine the extent of *B*-*C* ordering in the copper matrix, first-nearest neighbor (1NN) *B*-*C* ordering energies in Cu are calculated from the pairwise binding energies in Cu for the dilute regime shown in Equation (2) using the convention that negative ordering energies indicate preference for unlike pairs (i.e., *B*-*C*).

$$E_{\text{ordering}}(B\text{ - }C) = \frac{1}{2}[E_{\text{b}}(B\text{ - }B) + E_{\text{b}}(C\text{ - }C)] - E_{\text{b}}(B\text{ - }C) \quad (2)$$

where each $E_{\text{b}}$ is the binding energy for 1NN B-B, C-C, and B-C pairs, respectively.

### B. KineCluE Transport Coefficient Calculations:

Diffusion coefficients in the dilute ternary alloy are obtained as a sum over cluster contributions, each individually being computed using the open-source code KineCluE [4]. The kinetic properties of the binary alloy (e.g., V-B clusters) are described within the framework of the five-frequency model due to its simplicity in deriving analytical expressions for diffusion [39,40]. For the ternary alloy (e.g., V-B-C clusters), the kinetic properties are initially based on the kinetically resolved approximation (KRA), where the migration energy $Q^{\alpha}$ is solely determined by the migrating atom type (α), the initial configuration energy ($E_{\mathrm{b}}^{\mathrm{i}}$), and the final configuration energy ($E_{\mathrm{b}}^{\mathrm{f}}$) as shown in Equation (3) [41].

$$E_{\mathrm{mi}}^{ij} = Q^{\alpha} + \frac{E_{\mathrm{b}}^{\mathrm{i}} - E_{\mathrm{b}}^{\mathrm{f}}}{2} \tag{3}$$

Subsequently, a sensitivity analysis is performed with KineCluE [4] to identify which vacancy migration energies within a triplet cluster (V-*B*-*C*) most significantly influence the relevant transport properties of the alloy ($L_{BV}$, $L_{VV}$, and $L_{BB}$). The most sensitive migration energies are then accurately calculated using DFT. Once the transport coefficients have converged with respect to the subsequent DFT-calculated migration barrier, the final transport properties of the targeted ternary alloy are obtained. More details for this sensitivity analysis are included in the S.M. [35]. The kinetic range in KineCluE, e.g., the cutoff distance for effective interactions, is set to 4.0*a* (*a* is the matrix lattice parameter) for the pair and triplet clusters based on the convergence of the drag ratio $L_{ZrV}/L_{VV}$ and the cluster mobility of the V-Zr-Fe cluster (see Figure S2 in the Supplemental Materials). The thermodynamic range, e.g., the cutoff distance for thermodynamic interactions, was set to the 1NN shell based on a comparative analysis of 1NN and 5NN thermodynamic range calculations carried out for the V-Zr and V-Zr-Fe clusters (see Figure S1 in the S.M. [35]). These results indicate that the 1NN thermodynamic range is sufficient for elucidating trends associated with both two-body and three-body clusters in the Cu(Zr,Fe) and Cu(Zr,Co) alloys. With these parameters, there are 2 configurations and 5 jump frequencies for the V-B clusters (e.g., the five jump frequency model), which are all calculated using DFT in Table IV. For the V-*B*-*C* clusters, this kinetic and thermodynamic range results in 14 configurations and 53 jump frequencies, which shows the need for the sensitivity study in KineCluE. The triplet binding energies with 1NN Zr-*C* and 1NN Zr-V pairs are calculated using DFT, see section 3 in S.M. [35]. The sensitivity analysis indicates that it is sufficient to estimate the binding energies for the remaining configurations using additive pairwise binding energies. Similarly, the two most significant jump frequencies towards the transport coefficients of the three-body clusters are identified using KineCluE and calculated using DFT. The remaining jump frequencies are estimated with the KRA approximation.

In the ternary Cu(*B*,*C*) alloy, the driving forces controlling diffusion, within the framework of the thermodynamics of irreversible processes in the linear response regime [42], are the chemical potential gradients of V (vacancies), *B*, and *C* solutes, as shown in Equation (4), generating V, *B*, and *C* fluxes.

$$\begin{pmatrix} J_V \\ J_B \\ J_C \end{pmatrix} = -\begin{pmatrix} L_{VV} & L_{VB} & L_{VC} \\ L_{VB} & L_{BB} & L_{BC} \\ L_{VC} & L_{BC} & L_{CC} \end{pmatrix} \begin{pmatrix} \frac{\nabla(\mu_V - \mu_M)}{k_B T} \\ \frac{\nabla(\mu_B - \mu_M)}{k_B T} \\ \frac{\nabla(\mu_C - \mu_M)}{k_B T} \end{pmatrix} \quad (4)$$

where, the chemical potential of species i is $\mu_i$, and the Boltzmann constant is $k_B$. The flux of the matrix atoms (Cu) is such that the sum of all fluxes amounts to zero in vacancy-mediated diffusion. The number of transport, or Onsager, coefficients required to fully describe the flux of species resulting from an arbitrary set of driving forces in an isotropic and isothermal alloy containing *n* defects or solutes is $n(n + 1)/2$, owing to the symmetry of the Onsager matrix, i.e., 6 in the present study.

Since the solutes of interest do not form mixed-dumbbell interstitials in FCC Cu (see Appendix A1), solute transport is only mediated by vacancies. The total Onsager transport coefficients of the ternary alloy, Cu(*B,C*), are obtained by summing all the cluster contributions in the dilute limit as shown in Equation (5).

$$\begin{bmatrix} L_{VV} & L_{VB} & L_{VC} \\ L_{VB} & L_{BB} & L_{BC} \\ L_{VC} & L_{BC} & L_{CC} \end{bmatrix} = [V] \begin{bmatrix} L_{VV}(V) & 0 & 0 \\ 0 & 0 & 0 \\ 0 & 0 & 0 \end{bmatrix} + [VB] \begin{bmatrix} L_{VV}(VB) & L_B(VB) & 0 \\ L_{VB}(VB) & L_{BB}(VB) & 0 \\ 0 & 0 & 0 \end{bmatrix}$$
$$+ [VC] \begin{bmatrix} L_{VV}(VC) & 0 & L_{VC}(VC) \\ 0 & 0 & 0 \\ L_{VC}(VC) & 0 & L_{CC}(VC) \end{bmatrix} + [VBC] \begin{bmatrix} L_{VV}(VBC) & L_{VB}(VBC) & L_{VC}(VBC) \\ L_{VB}(VBC) & L_{BB}(VBC) & L_{BC}(VBC) \\ L_{VC}(VBC) & L_{BC}(VBC) & L_{CC}(VBC) \end{bmatrix} \quad (5)$$

where [V] is the concentration of vacancies, [VB] is the concentration of 1NN V-B pairs, [VBC] is the concentration of V-B-C triplets in any configuration with 1NN binding between pairs.

In this work, up to three-body clusters are considered to identify trends associated with triplet clustering. Currently, clusters larger than 4-body cannot be calculated using KineCluE with a sufficient kinetic range (e.g., $a_{kin}$=3.0) due to computational limitations as the number of configurations and jump frequencies grow exponentially with cluster size (see Figure S3). In KineCluE, local equilibrium is assumed between configurations of a given cluster. We further assume here that this local equilibrium condition also holds between clusters, meaning that the relative probabilities of isolated species (defect or solute), pair and 3-body clusters follow a Boltzmann equilibrium distribution. Cluster concentrations are then obtained using cluster partition functions $z_{Cl}$ given by Eq 6 for any cluster *Cl* and computed by KineCluE, under the constraint of fixed solute and defect nominal concentration.

$$z_{Cl} = \sum_{\rho \in \rho_{Cl}} g_\rho \exp\left(\frac{E_b(Cl_\rho)}{k_B T}\right) \quad (6)$$

where ρ is a single configuration of that cluster and $g_\rho$ is the geometric multiplicity of that cluster configuration. The transport coefficients of each cluster are obtained from the self-consistent mean field (SCMF) theory, which depends on the probability that a given jump occurs, expressed as $p_i\omega_{ij}$, where $p_i$ is the probability of being in configuration and $\omega_{ij}$ is the transition frequency from configuration *i* to configuration *j* [43]. Detailed balance at equilibrium leads to $p_i\omega_{ij} = p_j\omega_{ji}$. From transition-state theory, these jump rates are expressed using the Arrhenius law $\omega_{ij}=\nu_i \exp(-E_{mi}(ij)/k_B T)$, where $\nu_i$ is the attempt frequency and $E_{mi}(ij)$ is the migration energy, $k_B$ is the Boltzmann constant, and T is the absolute temperature.

Furthermore, cluster mobilities are obtained from transport coefficient calculations under the constraint that dissociative jumps (e.g., jumps that take the vacancy to a distance beyond the kinetic range from the solute) are prevented [4,44].

Two distinct environments are considered in this work for calculating the concentrations of clusters for the total alloy transport coefficients – the first one for alloys at equilibrium conditions, making use of the thermal concentration of vacancies and related equations for vacancy and solute diffusion listed in Appendix A1 from T. Schuler et al.'s work [17,28], and the second one for irradiation conditions considering steady-state vacancy concentrations under irradiation based on the homogeneous rate equations also listed in Appendix A1 following L. Huang et al.'s work [36]. The irradiation flux considered in this work is $1\text{x}10^{-3}$ dpa/s. The total sink strength ($k^2$) is set to $10^{15}$ $m^{-2}$ corresponding to an approximate grain size of 250 nm which will have a strong sink regime [44]. Gekko python package was utilized for solving the nonlinear system of equations for the concentrations of defects and solutes in the ternary alloys [45]. These concentrations are then used to calculate the total transport coefficients $L_{ij}$ defined in Eq. (5). Three quantities are employed to quantify the benefits brought about by the alloying element *C*, namely, the total vacancy diffusion coefficient, $\overline{D_V}$, defined as $\overline{D_V} = L_{VV}/[\bar{V}]$, where $[\bar{V}]$ is the concentration of free and bound vacancies; the total solute B diffusion coefficient, $\overline{D_B}$, defined as $\overline{D_B} = L_{BB}/[\bar{B}]$, where $[\bar{B}]$ is the concentration of free and bound *B* solutes; and the solute drag ratio, defined as $L_{BV}/L_{VV}$.

Under nonequilibrium steady state conditions (NESS) like under particle irradiation, the concentration is calculated using standard rate equation [35, 43] yielding Equation 7:

$$[\bar{V}]^{\mathrm{NESS}} = [\bar{V}]^{\mathrm{eq}} - \frac{k^2\Omega}{8\pi r_c} + \sqrt{\left(\frac{k^2\Omega}{8\pi r_c}\right)^2 + \frac{\phi\Omega}{4\pi r_c \overline{D_V}}} \tag{7}$$

where $\phi$ is the irradiation flux, $\Omega$ is the atomic volume, $k^2$ is the total sink strength for vacancies, and $r_c$ is the recombination radius, i.e., the distance below which V and I spontaneously recombine. $r_c$ is assumed to be independent of solute concentration or temperature at a constant of 7 Å. The irradiation flux considered in this work is $1\text{x}10^{-3}$ dpa/s. The total sink strength ($k^2$) is set to $10^{15}$ $m^{-2}$ corresponding to an approximate grain size of 250 nm which will have a strong sink regime. This would test if the *C* solute can reduce the sink regime. The vacancy concentration under these conditions is a function of the vacancy diffusion coefficient (see Equation A4) which is numerically solved self-consistently with the solute concentrations. This is initially set to the $L_{VV}(VBC)$ since the [V*BC*] term dominates the numerator and dominator in Eq S4 due to the high triplet binding energies. Further details are provided in Appendix A1.

## III. RESULTS:

### A. Screening of B-type and C-type solutes in FCC copper

First-nearest neighbor (1NN) pairwise interactions between 21 solute pairs are computed, choosing one from the strong vacancy binding region (*B*-type solute) and one from the slow diffuser region (*C*-type Solute) in the solute diffusion map (Figure 1). Zr, Ge, and Sn are selected as *B*-type solutes, ensuring large V-*B* binding energies without introducing solutes known for embrittling Cu alloys. It should be noted that Sn and Ge have a higher equilibrium solubility limit in Cu of around 6.2 wt% and 9.7 wt% at 350°C, respectively, compared to less than 0.01 wt% for Zr, though the second solute will affect this solubility limit [46–48]. For *C*-type solutes, Co, Mo, Ni, Fe, Cr, W and Nb are chosen due to their slower diffusion in Cu. Interestingly all *B*-*C* solute combinations in the FCC copper matrix ended up with attractive binding energies ranging from 0.01 to 0.49 eV (see Figure 2). From this pairwise screening, Zr-Co, Zr-Mo, and Ge-Mo are found to have significant 1NN binding energy ($E_{binding}^{B-C}$>0.3eV) in the copper FCC matrix. Zr-Co and Zr-Fe are selected to investigate in detail the role of the Zr-*C* binding energy on Zr solute diffusion in the copper FCC matrix.

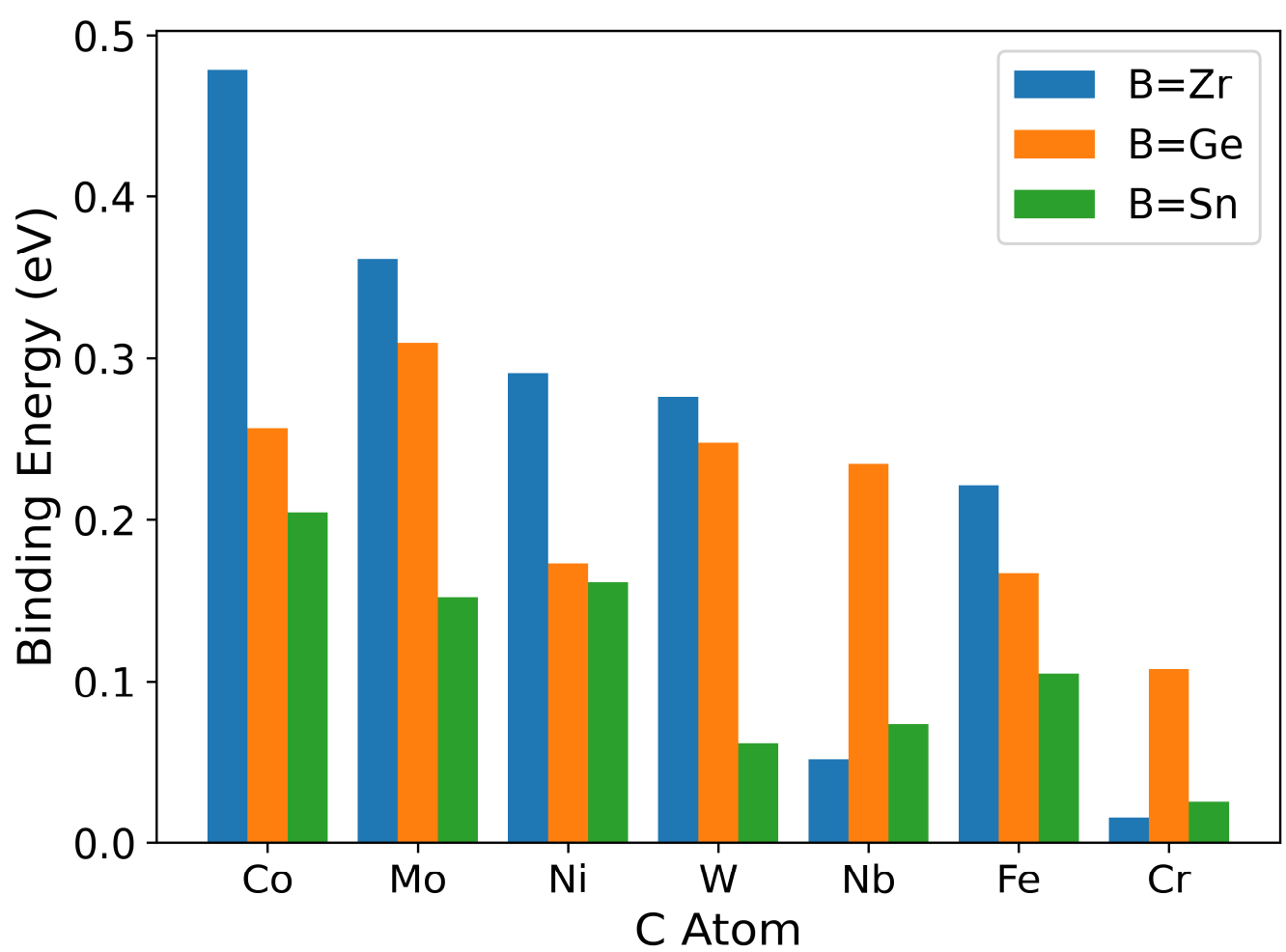

Figure 2: 1NN solute-solute binding energies between *B* and *C*-type solutes in FCC Cu matrix computed using DFT

### B. Thermodynamic Stability of Ternary V-Zr-Fe and V-Zr-Co Clusters in dilute Cu(Zr,Fe) and Cu(Zr,Co) alloys:

#### 1. Pairwise interactions with vacancies in dilute Cu(Zr,Fe) and Cu(Zr,Co) alloys

The pairwise binding energies between solutes and vacancies up to the 5th nearest neighbor (NN) distance in the Cu FCC matrix are calculated using DFT to identify the range of solute-solute and solute-defect interactions in the ternary alloys, as shown in Figure 3. In the Cu(Zr,Fe) alloy, the Zr-Fe binding energy at 1NN is 0.21 eV, which is comparable to the V-Zr binding of 0.25 eV but lower than the Fe-Fe binding of 0.32 eV. The Zr-Zr pairs are strongly repulsive with a 1NN binding energy of -0.274 eV.

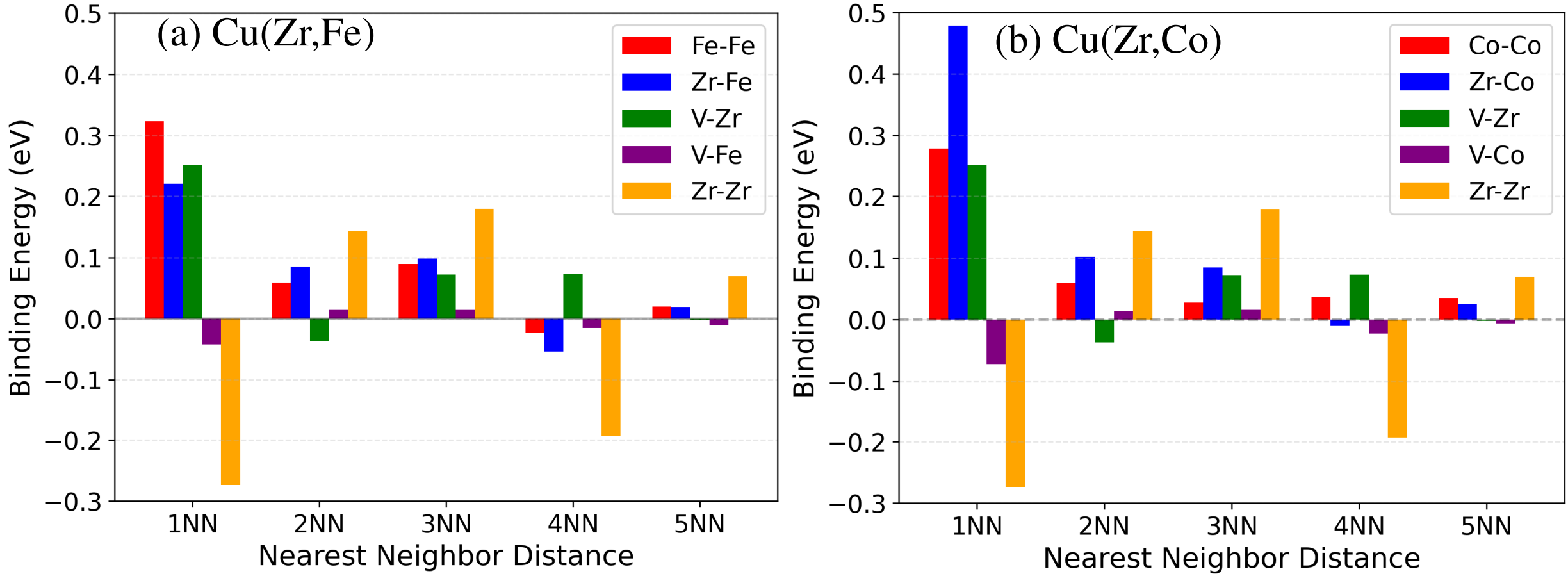

Figure 3: Pairwise binding energies up to 5NN in (a) Cu(Zr,Fe) and (b) Cu(Zr,Co) vacancy-solute clusters in FCC Cu, computed using DFT.

In contrast, the Cu(Zr,Co) alloy exhibits a much stronger Zr-Co binding energy of 0.48 eV at 1NN. This 1NN attraction is significantly larger than both the Zr-Vac binding (0.25 eV) and the Co-Co binding (0.27 eV), indicating a strong preference for the formation of Zr-Co pairs. This strong ordering tendency is further quantified by the ordering energies, which show a more negative ordering for Zr-Co (-0.478 eV) than for Zr-Fe (-0.196 eV). This preference for Zr-*C* ordering, driven by the strong Zr-C binding energy and repulsive Zr-Zr interactions, suggests that, in the ternary alloys, Zr may be primarily transported as part of V-Zr-*C* clusters rather than simple V-Zr pairs. Additionally, significant Zr-Fe and Zr-Co 1NN ordering can be expected within the FCC Cu matrix.

Interestingly, the presence of solutes and vacancies affects the magnetic properties of Fe and Co solutes. As shown in Table V, pairing with a vacancy at the 1NN position slightly increases the magnetic moment of Fe and Co, while pairing with Zr at the 1NN position significantly lowers their magnetic moments. The magnetic moment of the Fe solute in Cu agrees well with the value calculated by Eisenbach et al. [49].

### 2. Triplet Interactions for V-Zr-Fe and V-Zr-Co clusters in Cu

Given that Zr is expected to be transported within V-Zr-*C* clusters due to high Zr-*C* ordering energies, the three-body or "triplet" interactions are quantified in Table I to determine if the vacancy attraction is significantly altered by the presence of the second solute. The V-Zr-*C* triplet binding energies are calculated using DFT for configurations with 1NN Zr-*C* and 1NN Zr-V (see Figure C1 for configurations). These configurations are predicted to have the highest binding energies from the sum of the constituent pairwise interactions and therefore are likely to have the largest impact on the cluster transport coefficients. These triplet binding energies are then compared to the sum of the constituent pairwise interactions up to 5NN. The thermodynamic stability of the configurations is reduced for linear chain and bent chain configurations on (100) crystallographic planes, whereas it is enhanced for compact configurations on (111) planes when compared to the constituent pairwise interactions.

Table I: Binding energy of triplet V-Zr-Fe and V-Zr-Co clusters compared to binding energies expected solely from pairwise interactions up to 5NN, both calculated using DFT. See Figure C1 for configurations.

| | V-Zr-Fe Cluster | | V-Zr-Co Cluster | |
|---|---|---|---|---|
| Configuration (see Figure C1) | Triplet Binding Energy (eV/triplet) | Pairwise Binding Energy (eV/triplet) | Triplet Binding Energy (eV/triplet) | Pairwise Binding Energy (eV/triplet) |
| (100) triangle | 0.488 | 0.487 | 0.733 | 0.745 |
| (111) triangle | 0.453 | 0.431 | 0.776 | 0.659 |
| Bent chain | 0.433 | 0.487 | 0.609 | 0.746 |
| Linear chain | 0.423 | 0.458 | 0.571 | 0.708 |

To rationalize the strong triplet binding energies observed for triplet clusters in the (111) triangle configurations, the valence electron density distributions and interatomic distances are examined further, see Figure 4. These electron density maps reveal that the (111) triangle has increased electron density between the Zr and Co atoms when compared to the linear chain configuration. Comparing Figure 4(a) and 4(b) reveals that the reduced binding energy of the linear chain configuration is due to the Co solute pulling the Zr solute away from the vacancy. The Zr atom is also further away from the vacancy in the linear chain configuration signaling reduced vacancy binding energy to the Zr-Co pair. Analyzing similar electron density maps for V-Zr-Fe shows similar trends with the electron density significantly elevated between Zr and its neighboring Fe solute, relative to between Cu atoms of the host matrix in the (111) triangle configuration (approximately 0.08 e/Å³ for Zr-Co and 0.06 e/Å³ for Zr-Fe pairs, versus only 0.04 e/Å³ for Cu-Cu pairs). These higher electron densities suggest enhanced electronic interactions and stronger directional bonding in the Zr-Co pairs compared to Zr-Fe pairs. Furthermore, we note that the interatomic distances are consistent with the size mismatch of the solutes in the copper matrix, see section 4 of the S.M. [35].

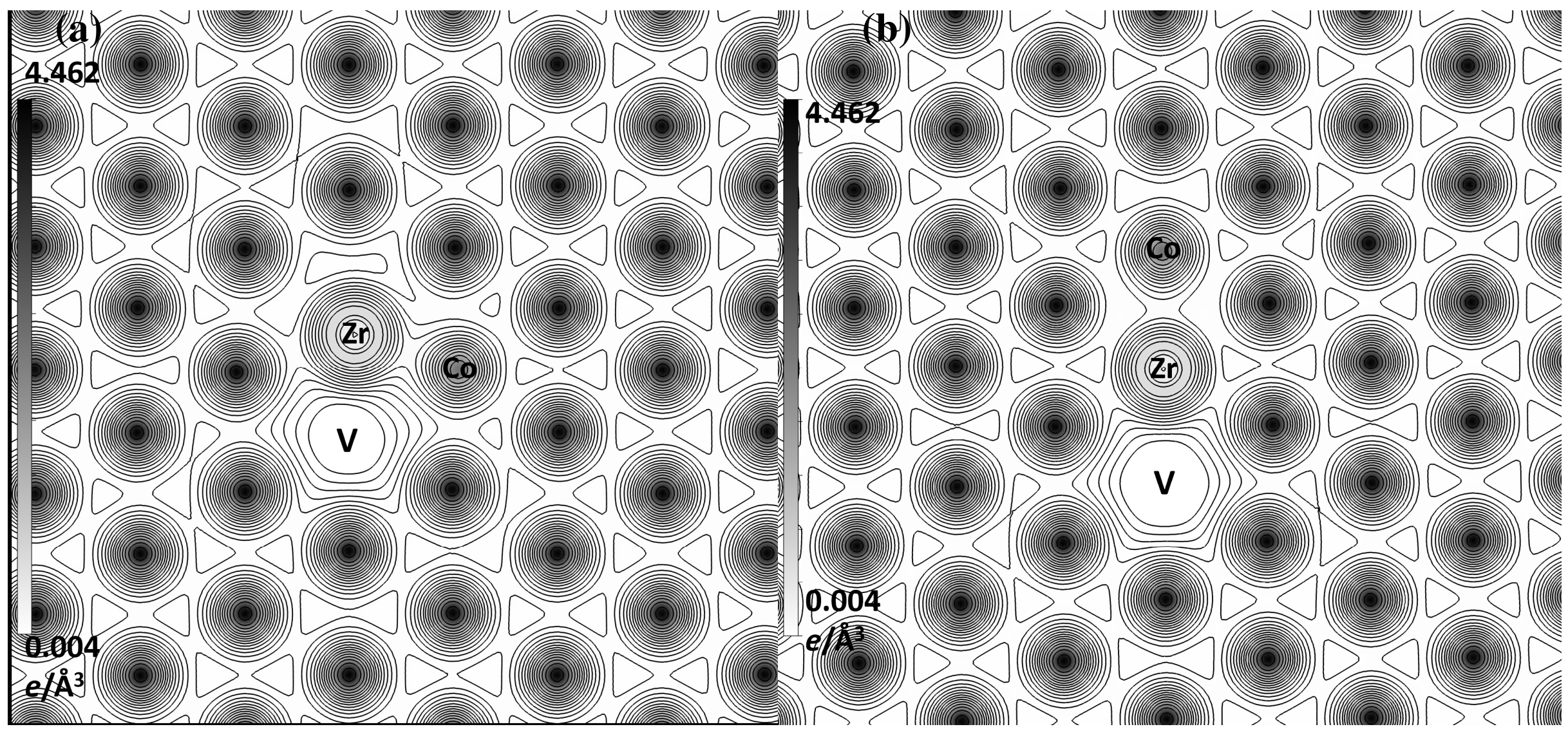

Figure 4: Valence electron density maps in the compact (111) triangle configuration in (a) and in the linear chain configuration in (b) for V-Zr-Co clusters in the FCC Cu matrix computed using DFT. A slice of the (111) plane is shown. Contour lines with logarithmic intervals are shown. Matrix Cu atoms are not labeled.

### C. Diffusion in Dilute Cu(Zr), Cu(Zr,Fe), and Cu(Zr,Co) alloys at thermal equilibrium

#### 1. Cluster Transport Coefficients for V-Zr, V-Zr-Fe and V-Zr-Co

The effects of a *C* solute (e.g., Fe or Co) on the mobility of the Zr solute are quantified by calculating the cluster transport coefficients, $L_{ij}$(cluster), for V-Zr, V-Zr-Fe, and V-Zr-Co clusters using KineCluE as shown in Figure 5. The introduction of either the Fe or Co solute significantly reduces the key transport coefficients in the relevant clusters, including $L_{ZrZr}$, $L_{VV}$, and $L_{VZr}$ , especially at lower temperatures. This reduction is even more pronounced in the V-Zr-Co cluster compared to the V-Zr-Fe cluster, which is a direct consequence of the stronger Zr-Co binding. For instance, the activation energy for the off-diagonal V-Zr coupling term, $L_{VZr,}$ is increased by 0.322 eV between the V-Zr to V-Zr-Co clusters, from 0.653 eV to 0.975 eV. This indicates that the increase in binding energy between Zr-C pairs can significantly reduce solute drag by altering the mobility of vacancies near the Zr-C pair. It is also interesting to note that the cluster transport coefficients show some curvature at higher temperatures, indicating that other diffusion paths are being activated.

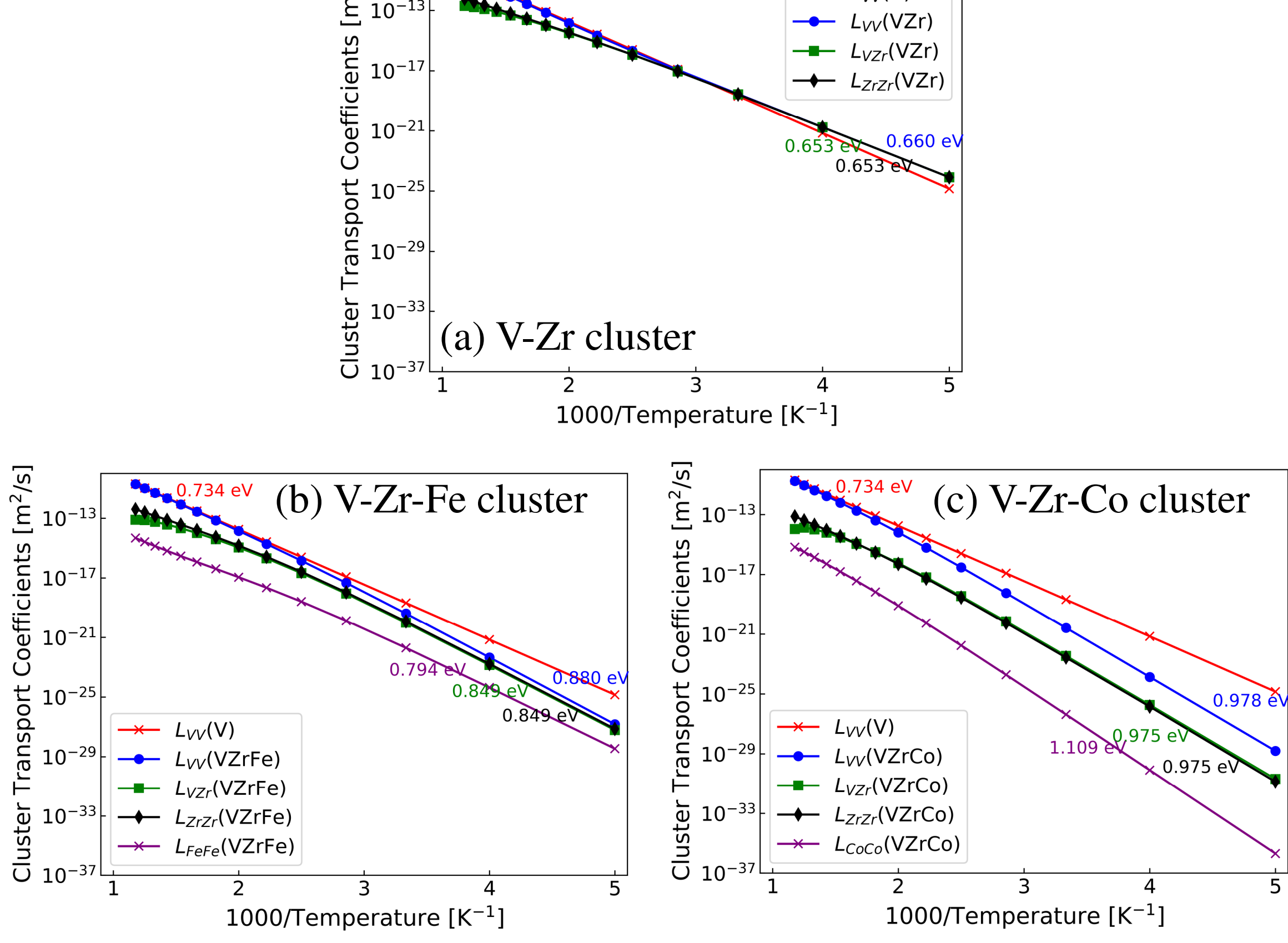

Figure 5: Cluster Onsager Transport Coefficients for V-Zr, V-Zr-Fe, and V-Zr-Co clusters with labeled activation energies calculated using KineCluE with 4a kinetic range and 1NN thermodynamic range.

**2. Solute and vacancy diffusion in Cu(Zr), Cu(Zr,Fe), and Cu(Zr,Co) alloys under equilibrium conditions**

Solute and vacancy diffusion coefficients under equilibrium conditions are tabulated as a sum of the cluster Onsager transport coefficients weighted by the cluster's concentration in the ternary alloys, as described in the Methods section and in Appendix A1-2. The concentrations of clusters are solved for each system and are compared in Figures A3. The diffusion coefficient of Zr is lowered in both the ternary Cu(Zr,Fe) and Cu(Zr,Co) alloys compared to the binary Cu(Zr) alloy across the entire temperature range studied in Figure 6. A comparison between the V-Zr-Fe and V-Zr-Co clusters reveals that Co is more effective at reducing the Zr mobility due to the high 1NN Zr-Co binding energy. For instance, the activation energy for Zr solute diffusion is raised from 1.574 eV to 1.821 eV, which is close to the activation energy for self-diffusion in Cu.

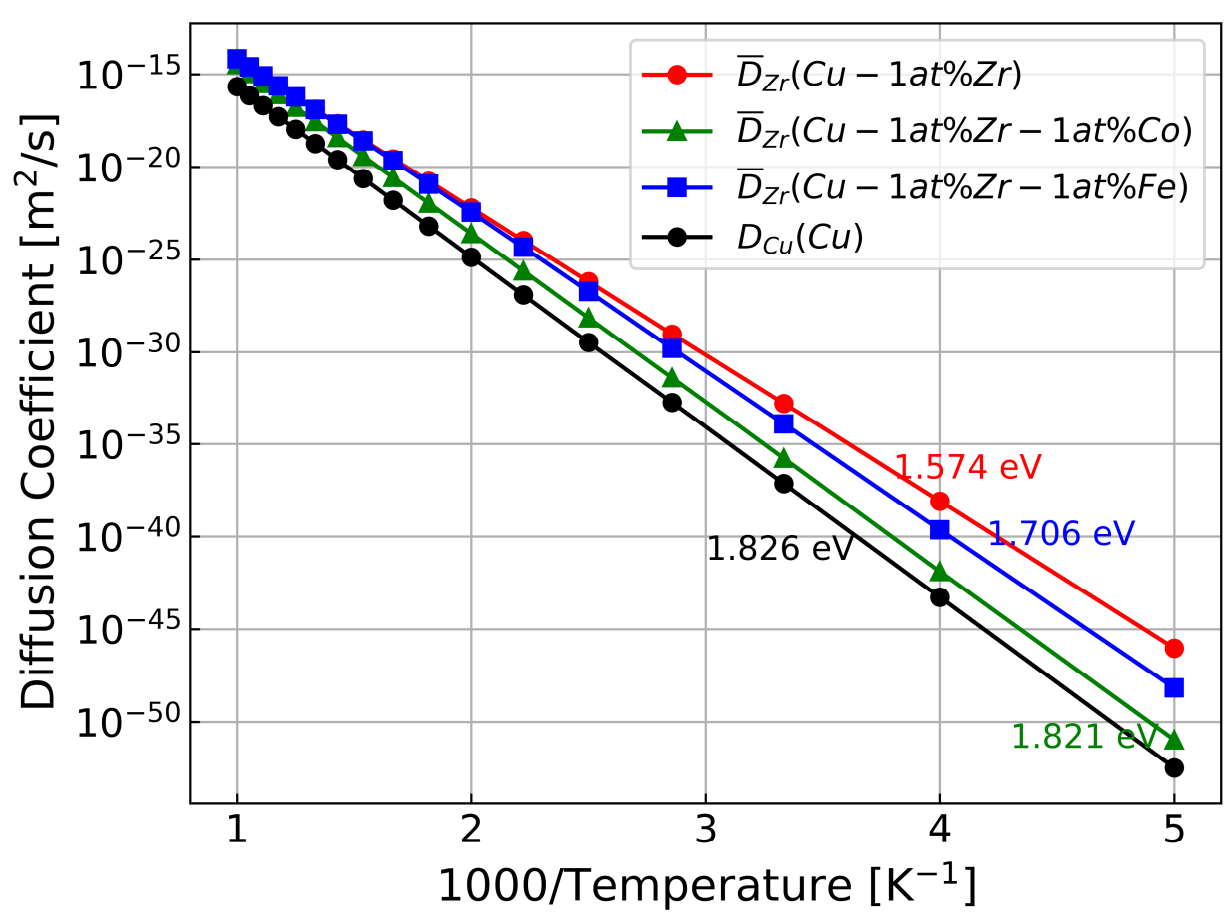

Figure 6: Equilibrium Zr solute diffusion in Cu(Zr), Cu(Zr,Fe), and Cu(Zr,Co) calculated at 1at% for each solute concentration alloys with labeled activation energies

Vacancy diffusion coefficients are then calculated for the dilute Cu(Zr), Cu(Zr,Fe), and Cu(Zr,Co) alloys, see Figure 7. It is evident that vacancy diffusion is significantly slowed down in the dilute ternary Cu(Zr,Co) and Cu(Zr,Fe) compared to the dilute binary Cu(Zr). In fact, the activation energy for vacancy diffusion is significantly increased by 0.318 eV from 0.660 eV to 0.978 eV solely due to the addition of V-Zr-Co clusters in the dilute Cu(Zr,Co) alloy. The vacancy diffusion in the Cu(Zr,Fe) alloy is reduced as well, however, not as significantly, indicating that the strong *B*-*C* binding energy is responsible for this reduction. It should be noted that, for the alloys of interest here, the vacancy concentration under irradiation is largely independent of the addition of the C solute due to the repulsive interaction between *C* and vacancies, see Figure A7 and Equation A4 in Appendix A. The reduction of vacancy diffusion in the ternary is therefore predominantly due to the high concentration and low mobility of the V-*B*-*C* clusters. Overall, these results indicate that the triplet clusters considered here are highly effective in slowing down vacancies and should thus enhance the point defect recombination rate over that of elimination at sinks.

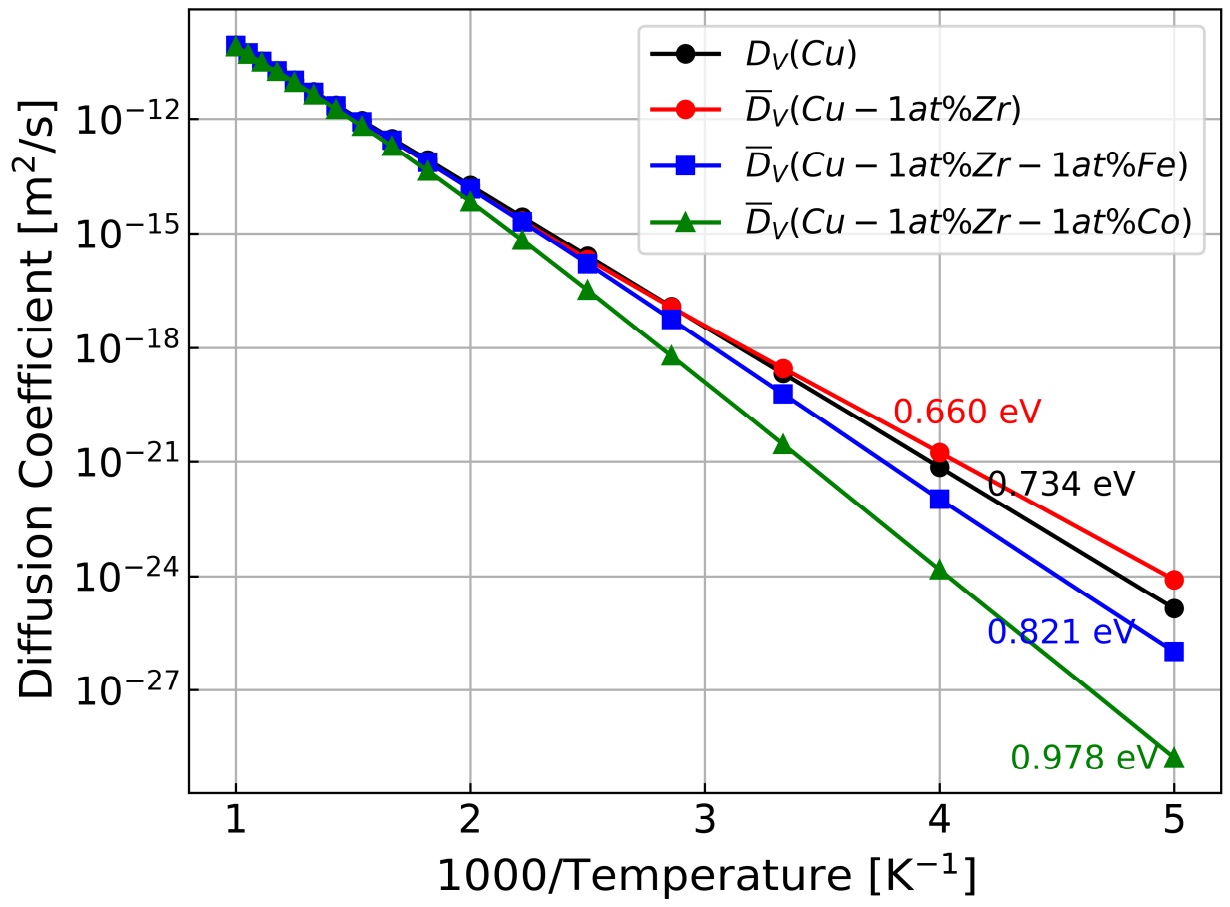

Figure 7**:** Vacancy diffusion in Cu(Zr), Cu(Zr,Fe), and Cu(Zr,Co) with labeled activation energies calculated at a nominal solute concentration of 1at% for each solute

Turning now to solute drag, we compare the vacancy-mediated solute drag ($L_{ZrV}/L_{VV}$) between these three alloys, see Figure 8, which shows that the Cu(Zr,Co) alloy experiences almost no solute drag compared to the other two alloys. The Cu(Zr,Fe) alloy showed about a 50% reduction in the solute drag at 500K to 0.09 compared to 0.19 for the Cu(Zr) alloy. For Cu(Zr,Co), this solute drag drops all the way to 0.009. This indicates that the V-Zr-Co clusters are highly effective at slowing down solute drag due to the high B-C binding energy.

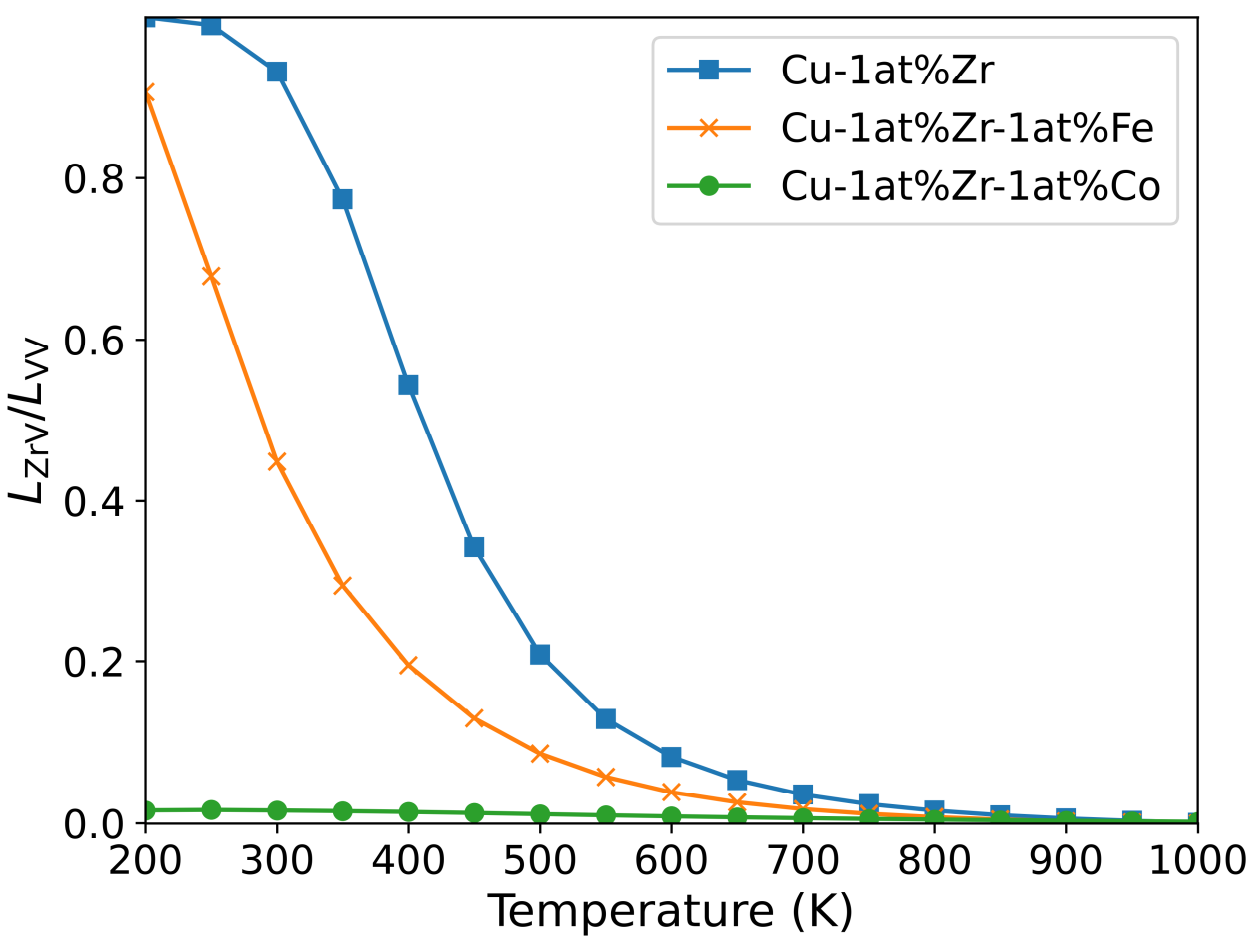

Figure 8**:** Vacancy-mediated solute drag ($L_{ZrV}/L_{VV}$) in Cu(Zr), Cu(Zr,Fe), and Cu(Zr,Co) alloys with a nominal solute concentration of 1at% for each solute

**D. Diffusion in Cu(Zr), Cu(Zr,Fe), and Cu(Zr,Co) alloys under irradiation**

Next, solute diffusion coefficients are calculated under irradiation conditions, when point defect concentrations have reached steady state (See Appendix A1 and A3. for more details). Under irradiation, the vacancy concentration is no longer solely determined by temperature but also by the production rate of defects and their elimination at sinks or through recombination. In pure metals, this leads to distinct temperature regimes for defect concentration: a high-temperature regime where thermal vacancies dominate, an intermediate sink-elimination regime where sustained vacancy fluxes to sinks can cause solute drag, and a low-temperature recombination-dominated regime. The ideal ternary alloy would promote recombination by extending the temperature range of the recombination regime, thereby reducing the sink regime temperature range. The concentrations of clusters under steady-state irradiation are numerically solved and are shown in Figures S7-9 for a temperature range of 300-1000K. The resulting solute diffusion plots for the same temperature range are shown in Figure 9.

The diffusion of Zr is slower in the ternary Cu(Zr,Fe) and Cu(Zr,Co) alloys compared to the binary Cu(Zr) alloy, as illustrated in Figure 9. The strong binding between Zr and the C solute traps Zr in less mobile V-Zr-C clusters, reducing the concentration of isolated Zr that is susceptible to being dragged by vacancies. Crucially, the addition of the Co solute completely suppresses the sink-elimination regime prominent in the binary Cu(Zr) alloy. This is visible in the diffusion plots for Zr under irradiation, see Figure 9(a), where there is a broad plateau from 600K to 750K, characteristic of the sink regime in the Cu(Zr) binary alloy (red line), which is completely reduced in the Cu(Zr,Co) ternary alloy (orange line). This demonstrates that the V-Zr-Co clusters are especially effective at mitigating the vacancy-solute drag effect for the Zr solute under irradiation by extending the recombination regime in the Cu(Zr,Co) ternary alloy. In contrast, the diffusion of the *C* solutes (Fe and Co) is faster in the ternary alloys than their respective binaries, which is consistent with the high concentration of V-*B*-*C* clusters compared to V-*C* clusters. However, it should be noted that Fe and Co solutes in the respective Cu(Zr,*C*) ternary alloys are relatively slow diffusers since they are slower than the self-diffusion in Cu, as shown in Figure 9(b-c).

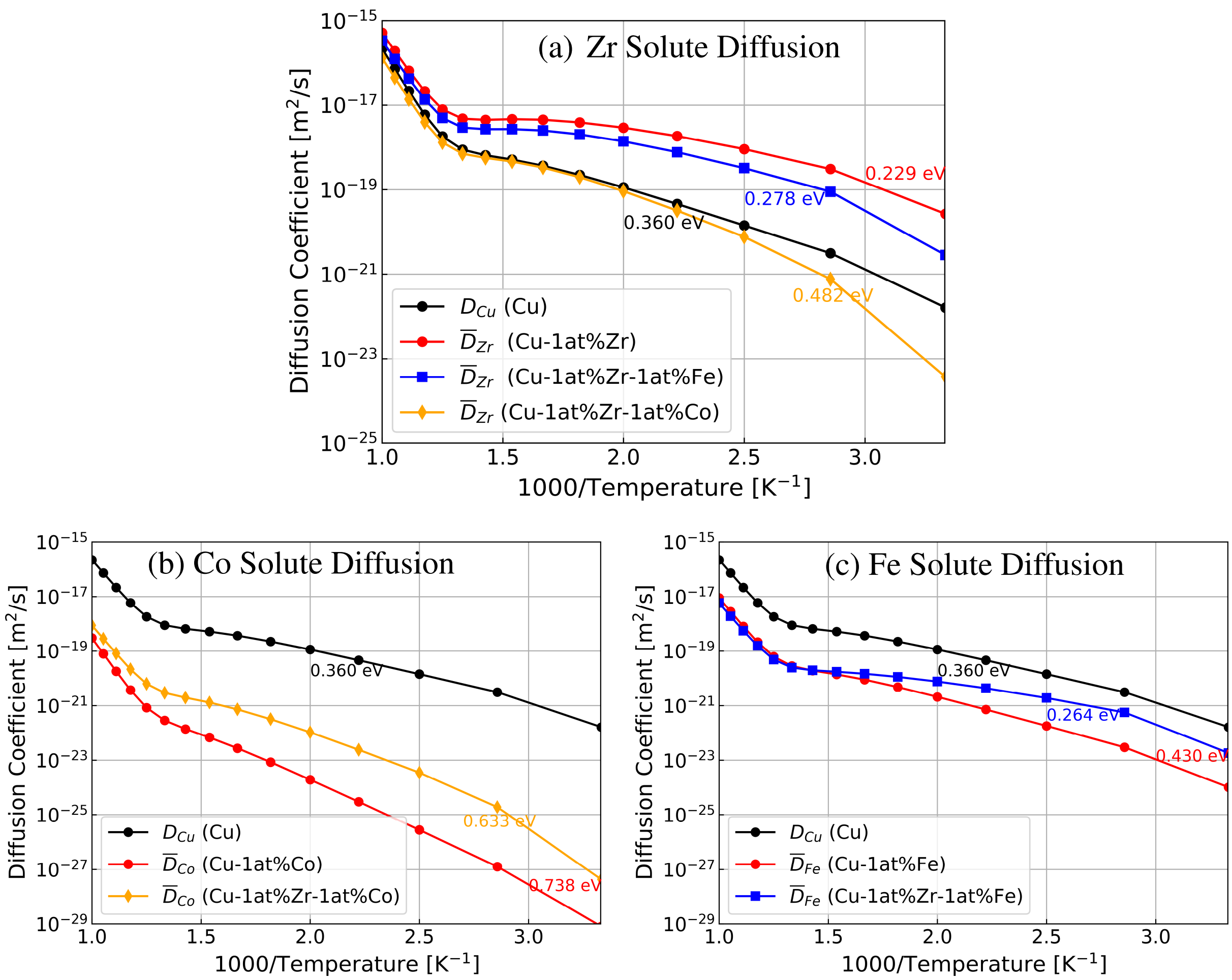

Figure 9**:** Solute diffusion coefficients calculated for Cu(Zr), Cu(Fe), Cu(Co), Cu(Zr,Fe), and Cu(Zr,Co) alloys under irradiation for vacancy and solute concentrations calculated with a total nominal solute concentration of 1at% for each solute. Zr diffusion in Cu(Zr,Co) and Cu(Zr,Fe) alloys is compared to binary Cu(Zr) in (a). In (b), Co diffusion is compared between Cu(Zr,Co) and Cu(Co) alloys, while in (c), Fe diffusion is compared between Cu(Zr,Fe) and Cu(Fe) alloys. The irradiation flux considered here is $1 \times 10^{-3}$ dpa/s. The total sink strength ($k^2$) is set to $10^{15}$ $m^{-2}$ corresponding to an approximate grain size of 250 nm which will have a strong sink regime.

## IV. DISCUSSION

A fundamental challenge in designing radiation-resistant dilute alloys is that B solutes chosen for their strong vacancy-binding properties are always fast diffusers in their respective binary alloys. It was recently proposed that adding a second, synergistic C solute could address this issue, and this approach is supported by a parametric KMC study using model ternary alloy systems [20]. In the present work, we have evaluated the potential of this alloy design strategy more quantitatively by calculating the role of the promising $C$ solutes in stabilizing dilute Cu-Zr-based alloys. As shown in the Results sections, Co as the $C$ solute is found to be the most effective at stabilizing V-Zr clusters, reducing diffusion of vacancies and suppressing Zr drag. In the present section we provide a perspective for some of the key findings of this work.

The reduction of the Zr solute diffusivity in ternary alloys is best appreciated by comparing it with other B solute diffusion in copper, see Figure 10. The inverse correlation between vacancy binding and solute diffusivity in dilute binary copper alloys is quantitatively established: This strong linear inverse correlation ($R^2$=0.92) with a slope of -1.03 demonstrates that all solutes with high vacancy-binding energies in binary copper alloys are also fast diffusers. When the vacancy binding energy is strong, it can be shown from Equation (C3) in the Appendix, that the activation energy for solute diffusion is approximately $Q_{SD} - E_B^{V-B}$, where $Q_{SD}$ is the activation energy for the self-diffusion of copper [50]. The excellent agreement between the observed slope (-1.03) and the theoretically predicted slope (-1) highlights the dilemma in designing radiation-resistant binary alloys: the very solutes that are best at trapping vacancies are also the most mobile ones, resulting in RIS and RIP in these alloys, and thus a degradation of their radiation resistance over time. This work demonstrates that introducing a second, synergistic C solute can overcome this problem. The ternary Cu(Zr,Co) alloy in Figure 10, marked by the "Zr-Co" label, shows the dramatic improvement derived from the addition of the Co solute, where the activation energy for Zr solute diffusion is raised by as much as 0.247 eV compared to the Cu(Zr) binary. Furthermore, it demonstrates that dilute Cu($B,C$) alloys are no longer constrained by the inverse correlation shown in Figure 10, opening up a new design space for radiation-resistant alloys.

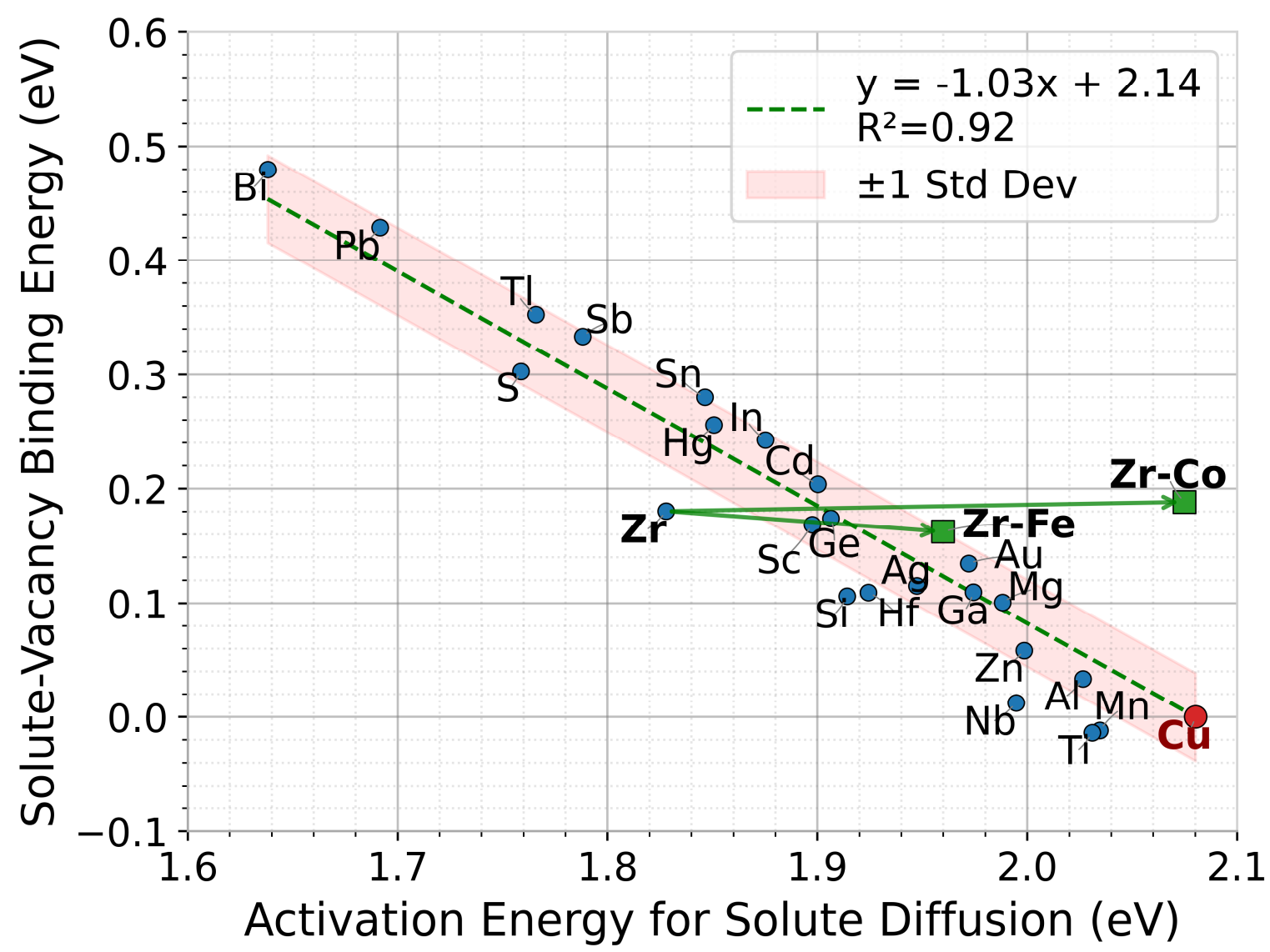

Figure 10**:** Activation energies for solute diffusion for the binary and ternary Cu alloys, for high vacancy binding energy solutes. Zr-Fe and Zr-Co labels (squares) indicate the Zr diffusion in Cu(Zr,Fe) and Cu(Zr,Co) alloys. The shift from the binary Cu(Zr) to ternary Cu(Zr,Fe) and Cu(Zr,Co) alloys is indicated by the two green arrows. The shifts to the right are due to the Zr-C binding energy. The shift downward for the Cu(Zr,Fe) alloy is due to the reduction in effective binding of vacancies (see Appendix C1).The data for binary alloys in this plot is from H. Wu et al. [27]. The activation energy corrections employed for binary alloys in ref. [28] is applied here as well for the ternary alloys, see Appendix A5 for more information.

In addition to the reduction of Zr solute diffusion, the Zr solute drag, defined by $L_{VZr}/L_{VV}$ (see Methods section), is significantly reduced between the Cu(Zr) and Cu(Zr,Co) alloys previously shown in Figure 8. Simplified but systematic KineCluE calculations are performed to quantify the effect of the Zr-C interactions on the Zr solute drag, by varying Zr-*C* binding energy and using KRA to estimate the migration energies. The resulting plot in Figure 11 shows that the Zr vacancy-mediated solute drag is reduced effectively by increasing the Zr-C binding energy to 0.5 eV. We find that the Cu(Zr,Co) alloy has almost no Zr vacancy-mediated solute drag compared to the binary Cu(Zr) at the temperatures of interest. This further demonstrates the importance of the strong B-C binding energy in screening these dilute ternary alloys. In this regard, Co is an ideal C solute since it completely suppresses the vacancy-mediated solute drag owing to the large binding energy with the Zr solute. Figure 11 reveals additionally that with large Zr-*C* binding energies, the C diffusivity in the Cu matrix is not an important factor controlling the Zr solute drag, in contrast to initial design criteria, see Introduction. A beneficial consequence is that more mobile *C* solute than initially anticipated, such as Nb or Al, could facilitate the seeding of the initial microstructure with solute clusters prior to irradiation [20].

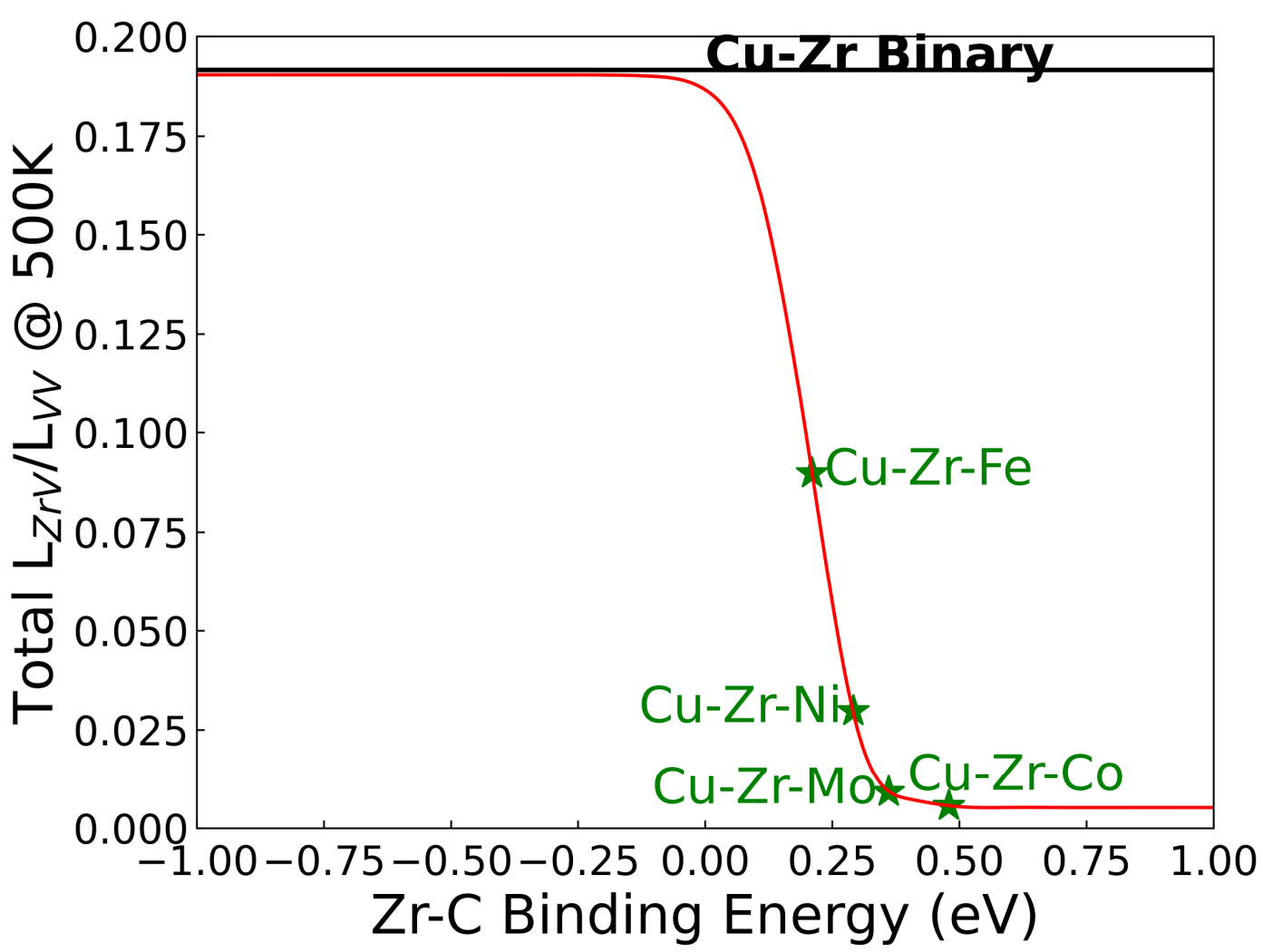

Figure 11**:** Effect of B-C binding energy on $L_{VZr}/L_{VV}$ calculated using KineCluE with kinetic range of $4a_0$ and thermodynamic range of 1NN with the location of several dilute ternary Cu(Zr,*C*) alloys shown on this curve

The present work also provides a mechanistic understanding of the reduction of solute drag by the addition of a second solute. For the dilute binary Cu(Zr) alloy, vacancy-mediated solute drag arises from the interplay between distinct vacancy jumps around the Zr solute, see Appendix B1. The low $E_3^m$ and $E_4^m$ activation energies compared to $E_1^m$, in this dilute alloy, enable rapid dissociation and reassociation of the vacancy with various first nearest neighbor (1NN) positions of the Zr solute, resulting in an effective rotational motion of the vacancy around the Zr atom. Consequently, this mechanism induces solute drag despite the high $E_1^m$ barrier, see additional details in Appendix B1. This contrasts, for example, with the origin of solute drag in other alloys like some binary Al alloys where the $E_1^m$ is lower than $E_3^m$ causing vacancy-rotations around the solute to be preferable to vacancy dissociation-reassociation [18]. In the ternary cases of dilute Cu(Zr,Fe) and Cu(Zr,Co) alloys, the results reported here show that the diffusion circuits of the vacancy are strongly affected by the presence of the *C* solute since the strong binding energy between Zr and *C* guarantees that transport is mediated by V-Zr-*C* clusters instead of simple V-Zr pairs. Specifically, the Zr-*C* pairwise interaction inhibits the kinetics of the vacancy's reassociation with sites occupied by the *C* solute, but without significantly modifying the V-Zr binding energy. Collectively, this results in the lower Onsager transport coefficients of vacancies, Zr solute, and *C* solute in these clusters compared to V-Zr clusters. The extent of this reduction depends on the strength of the *B*-*C* binding energy.

The improvements introduced by the second solute can be directly visualized by mapping Zr solute drag ratios for each dilute ternary alloy with effective migration energies for a fictitious, effective dilute binary alloy, Cu($B_{\text{eff}}$). To accomplish this, we overlay the ternary Zr solute drag ratio, $L_{VZr}/L_{VV}$, onto the binary solute drag ratio heatmaps calculated using the five-frequency model, see Figure 12. Here, the ratio of $L_{BV}/L_{VV}$ gives the tendency of a vacancy to carry a solute *B* (e.g., Zr) to a point-defect sink under irradiation. Starting with the five-frequency framework, there are five unique energies controlling the diffusion of solute and point defects: $E_0^m$, $E_1^m$, $E_2^m$, $E_3^m$, and $E_{binding}^{B-V}$. The bulk migration energy of vacancies in Cu is $E_0^m$ which is fixed at 0.734 eV. For $E_{binding}^{B-V}$, an effective binding energy to vacancies has been calculated for the Zr-Fe and Zr-Co solute pairs using DFT calculations, see Figure C1. The Co solute does not significantly affect the vacancy binding energy to Zr since the effective binding energy is increased by only 9 meV at 500K. Neglecting this small effect, effective $E_1^m$, $E_2^m$, and $E_3^m$ can now be directly estimated for the Cu(Zr,Co) alloy from the $L_{BV}/L_{VV}$ heatmaps in Figure 12 which are calculated using the five-frequency framework at a fixed $E_{binding}^{B-V}$. From this heatmap, it is evident that the dilute binary Cu(Zr) alloy is in a regime where slight changes in $E_1^m$ would not impact the solute drag ratio. This is supported by directly calculating the migration energy for various $E_1^m$-type jumps for Cu(Zr,Co) using DFT, see Figure C4. Therefore, the reduction in $L_{BV}/L_{VV}$ must come from an increase in the effective $E_3^m$ by approximately 0.1 eV, thereby shifting the Cu(Zr,Co) alloy to the left on the heatmap. For the Cu(Zr,Fe) alloy, however, a reduction in the effective binding energy of around 18 meV at 500K is observed for V-Zr-Fe clusters (see Appendix C1). This would result in a shift in the $L_{BV}/L_{VV}$ heatmaps (see Appendix C2). Ultimately, half of the decrease in drag ratio (~0.05 in $L_{ZrV}/L_{VV}$) is a result of the decrease in effective binding energy for the V-Zr-Fe clusters. Altogether, this analysis provides a simplified physical picture: the origin of the reduction in *B* solute drag by the addition of *C* solute is due to an increase in the activation barrier for vacancy dissociation from the effective *B* solute.

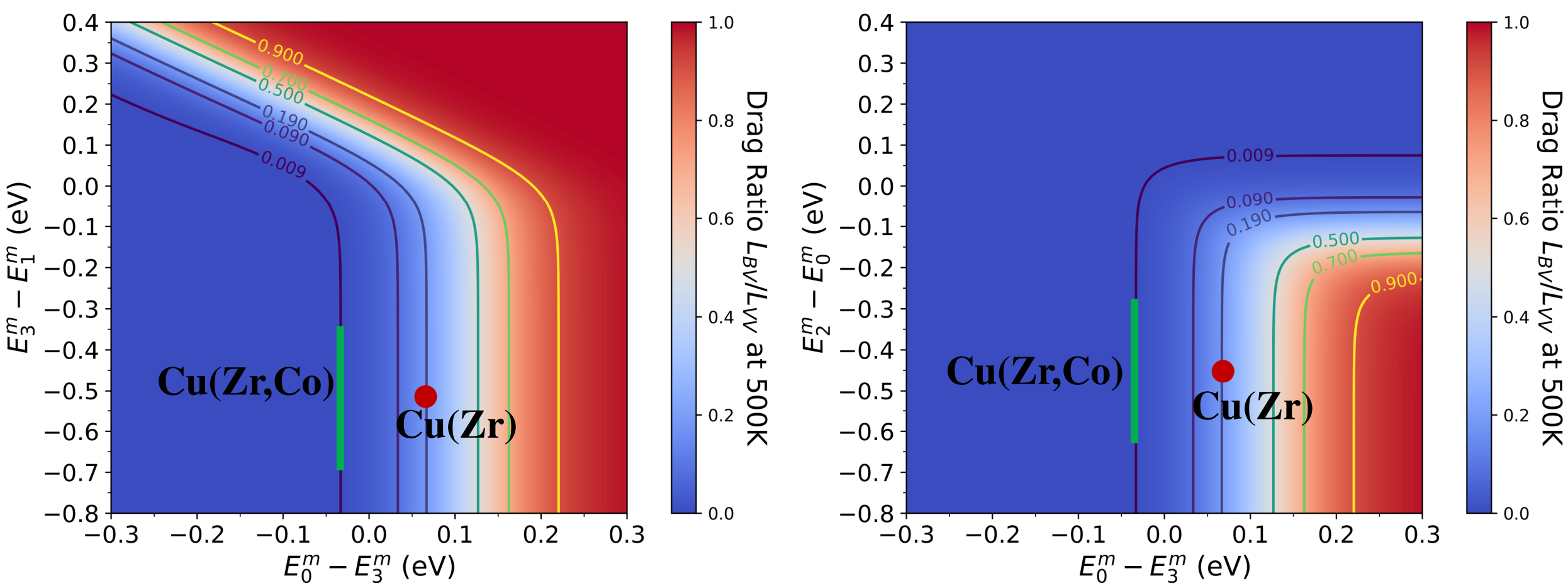

Figure 12**:** Solute drag maps ($L_{BV}/L_{VV}$) showing location of binary Cu(Zr) and ternary Cu(Zr,Co) indicating shift in effective $E_3^m$ between the binary and ternary alloys calculated using KineCluE with fixed $E_{binding}^{B-V} = 0.181\ eV$ at 500K.

The results here, thus far, have shown the impact of the *C* solute through triplet V-*B*-*C* clusters. Dilute alloy alloys may contain many-atom solute clusters larger than triplets, so a first attempt at assessing the effects of larger cluster sizes on cluster mobility is performed by extending calculations to include V-Zr-Fe-Fe and V-Zr-Co-Co clusters, for which high thermal stabilities are predicted given the strong constituent pairwise attractions. High cluster mobility generally leads to solute drag due to solute and point defects diffusing together to sinks. The 4-body cluster mobilities in Figure 13 are calculated in KineCluE using pairwise thermodynamic binding and the KRA model for jump frequencies for the sake of computational efficiency. It is evident that the cluster mobility significantly decreases as cluster size increases from V-Zr to V-Zr-Fe and to V-Zr-Fe-Fe clusters (red lines in Figure 13). A stronger reduction can be seen from V-Zr to V-Zr-Co-Co cluster mobilities in the Cu(Zr,Co) alloy (blue lines in Figure 13). These trends indicate that when *C-C* binding is significant as with Fe-Fe and Co-Co pairs, the mobility of higher-order clusters can be reduced with additional *C* solute. In certain 4-body configurations, additionally, there are two *B-C* interactions leading to a stronger cluster thermodynamic stability. Ultimately, the overall effect on diffusion and solute drag for each ternary alloy depends on the cluster mobilities and concentration of each cluster size under irradiation which in turn depend on the thermodynamic stability of each cluster size as well as their atomistic jump frequencies. Under irradiation, these effects become more complex due to the addition of radiation-enchanced diffusion and ballistic mixing as discussed next.

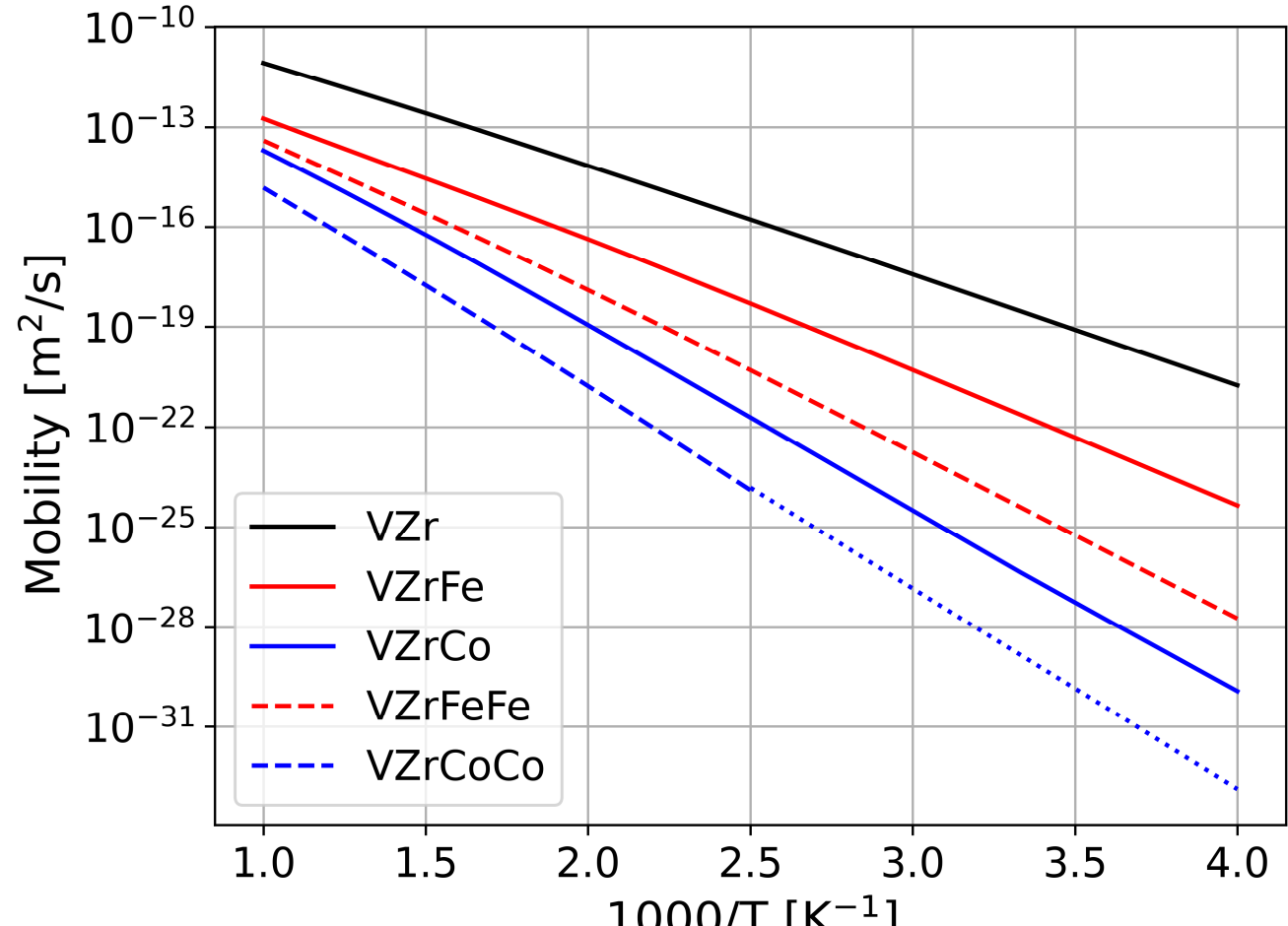

Figure 13**:** Cluster mobilities for VZr, VZrFe, VZrFeFe, VZrCo, and VZrCoCo clusters calculated using KineCluE with kinetic range of 2a and thermodynamic range of 1NN with pairwise binding energies for 3 and 4-body clusters and KRA model for jump frequencies. Due to a numerical instability at low temperatures for the V-Zr-Co-Co cluster calculation, a linear extrapolation is included from 250K-400K indicated by the dotted line.

While the above results provide novel insights into the 3-body thermodynamic and kinetic interactions of vacancies and solutes within dilute ternary copper-based alloys, there are several ways to extend this work. The present calculations do not include the ballistic mixing of chemical species forced by irradiation or the possible precipitation of solutes *B* and *C*, given their limited solubility and their attractive interaction. Fortunately, solute solubility can be greatly enhanced by irradiation, as demonstrated experimentally for a series of phase-separating Cu-based binary alloys [51]. Moreover, modeling and atomistic simulations have shown that ballistic mixing and point-defect recombination can also prevent clusters from coarsening [36,52]. The evolution of the cluster population under irradiation in specific Cu*(B,C*) systems could be investigated in future works using kinetic Monte Carlo simulations, extending the work done by S. Jana et al. [20] on model ternary alloys. Reactor pressure vessel steels may constitute a good example here, since they contain solute-defect clusters with more than 2 solute species such as MnSiCu clusters. [53]

The evolution and migration of these clusters under irradiation could be modeled, for instance, using lattice Kinetic Monte Carlo simulations since KineCluE code cannot handle large cluster sizes. For these KMC simulations, the binding energies and migration energies for vacancies near these solute clusters could be parametrized from DFT calculations. One strategy would be to develop machine learning potentials from DFT calculations to predict unknown binding energies and migration energies.

The main results in the present work are likely to be applicable to vacancy-mediated solute drag in other alloys such as aluminum or nickel-based alloys. For example, in Al and Ni matrices, the same correlation between strong vacancy-solute binding solutes and fast solute diffusion is also observed. Ternary aluminum alloys would be of particular interest given that commercial age-hardening aluminum alloys are based on solute additions with large binding energies, for instance Mg-Si in 6XXX and Mg-Zn in 7XXX [21]. Consequently, the methodology used here could be extended to evaluate solute pairs in these alloy alloys that would mitigate solute drag for these types of solutes.

Lastly, it will be important to experimentally validate the present computational predictions. For instance, thin copper alloy films suitable for ion irradiation could be characterized by scanning transmission electron microscopy and atom probe tomography to quantify and correlate radiation-induced segregation and solute clustering after irradiation.

## V. CONCLUSION

A novel approach has been proposed recently for the design of radiation-resistant dilute alloys, by addition of two synergistic solute species, a first solute, B, that binds to vacancies, and a second solute, C, that binds to solute B and suppress the mobility of V-B clusters. These stable vacancy-solute clusters impart lasting radiation resistance by promoting interstitial-vacancy recombination. In this work, first-principles calculations and KineCluE transport coefficients calculations for vacancies, solute, and vacancy-solute clusters are employed to first screen suitable *B* and *C* solutes in a Cu matrix and then quantify the improvements in cluster stability and recombination rates. One promising *B* solute, Zr, combined with two promising *C* solutes, Co and Fe, have been studied in detail. A stronger Zr-Co binding (0.48 eV) compared to Zr-Fe binding (0.21 eV) in the copper FCC matrix results in a much more pronounced reduction in Zr diffusion and solute drag in the dilute Cu(Zr,Co) alloy. The strong *B*-*C* attraction was found to be a crucial parameter to promote the formation of thermodynamically stable V-Zr-*C* triplet clusters, which become the primary cluster responsible for Zr diffusion, instead of the more mobile V-Zr pairs. Due to this effect, the activation energy for Zr diffusion increases from 1.574 eV to 1.821 eV in the Cu(Zr,Co) alloy, approaching the rate of Cu self-diffusion. Vacancy diffusion is also significantly hindered, with the activation energy for vacancy migration increasing by 0.318 eV in the Cu(Zr,Co) alloy compared to the Cu(Zr) binary, which enhances the probability of vacancy-interstitial recombination. The Zr solute drag ratio, $L_{ZrV}/L_{VV}$, was highly suppressed for the Cu(Zr,Co) alloy when compared to C(Zr) and Cu(Zr,Fe) alloys. The kinetic origin of this effect was analyzed by mapping the complex ternary interactions onto an effective binary five-frequency model. The analysis revealed that the strong Zr-*C* interaction disrupts the vacancy-solute exchange pathways, significantly increasing the energy barrier for a vacancy to dissociate from a Zr atom ($E_3^m$). These 3-body V-*B*-*C* clusters thereby overcome the inverse correlation between vacancy-binding and solute diffusivity that is found in binary alloys, opening up a new design space for radiation-resistant solid solutions which maintain their increased recombination effect over time.

## APPENDIX A: CALCULATION OF DIFFUSION COEFFICIENTS

### 1. Methods for calculating total diffusion coefficients

Under equilibrium, the vacancy concentration is determined by the temperature and the binding energy of vacancies to solutes. The equilibrium concentration of isolated vacancies is estimated from the enthalpy of formation: $[\mathrm{V}]^{\mathrm{eq}} = \exp(-\mathrm{H_V}/\mathrm{kT})$. Then, the total concentration of vacancies in the presence of solute *B* and *C* is given by Equation A1 [17,18,25,28].

$$[\overline{\mathrm{V}}]^{eq} = [\mathrm{V}^{\mathrm{eq}}][(1 + \mathrm{Z_{VB}}[B] + \mathrm{Z_{VC}}[C] + \mathrm{Z_{VBC}}[B][C]] \tag{A1}$$

where the total concentrations are notated with an overline (i.e. $\overline{[\mathrm{X}]}$). The isolated solute concentrations are notated without the overline (i.e. [X]).

To calculate the concentration of isolated B and C solute, Equation A2 is used reflecting the effect of V-*B*, *B*-*C*, and V-*B*-*C* clusters on the isolated solute concentration. In this work, the total solute concentration is set to 1 at.%.

$$[B] = \frac{[\overline{\mathrm{B}}]}{[1 + \mathrm{Z_{VB}}[\mathrm{V}] + \mathrm{Z_{BC}}[C] + \mathrm{Z_{VBC}}[\mathrm{V}][C]]} \tag{A2}$$

Vacancy and solute diffusion coefficients are expressed as an average over cluster transport coefficients weighted by cluster concentrations as shown in Equations A3 and A4 [18,36]. When V-*C* binding is negative as is the case with most *C* type solute, see Figure 1, the [V*C*] terms become negligible. When V-*B*-*C* binding energies are large, [V*BC*] becomes the dominant term.

$$\overline{\mathrm{D_V}} = \frac{\mathrm{L_{VV}}}{[\overline{\mathrm{V}}]} = \frac{[\mathrm{V}]\mathrm{L_{VV}}(\mathrm{V}) + [\mathrm{VB}]\mathrm{L_{VV}}(\mathrm{VB}) + [\mathrm{VC}]\mathrm{L_{VV}}(\mathrm{VC}) + [\mathrm{VBC}]\mathrm{L_{VV}}(\mathrm{VBC})}{[\mathrm{V}] + [\mathrm{VB}] + [\mathrm{VC}] + [\mathrm{VBC}]} \tag{A3}$$

$$\overline{\mathrm{D_B}} = \frac{\mathrm{L_{BB}}}{[\overline{\mathrm{B}}]} = \frac{[\mathrm{V}]\mathrm{Z_{VB}}\mathrm{L_{BB}}(\mathrm{VB}) + [\mathrm{V}][\mathrm{C}]\mathrm{Z_{VBC}}\,\mathrm{L_{BB}}(\mathrm{VBC})}{1 + [\mathrm{V}]\mathrm{Z_{VB}} + [\mathrm{C}]\mathrm{Z_{BC}} + [\mathrm{V}][\mathrm{C}]\mathrm{Z_{VBC}}} \tag{A4}$$

Mixed-dumbbell interstitials for Fe, Zr and Co are found to be not stable in FCC Cu using DFT calculations as described in the Methods Section of the main manuscript. The Zr mixed-dumbbell interstitial would not relax geometrically due to a large Zr solute size (>1eV). The Fe mixed-dumbbell interstitial had an unfavorable binding energy of -0.285 eV. Therefore, these solutes can only diffuse via vacancies in copper.

### 2. Solved Concentration of Clusters under equilibrium conditions

The concentrations of isolated solute, pairs and triplet clusters are calculated by assuming the relative probabilities of clusters follows a Boltzmann equilibrium distribution constrained by the local nominal concentration of solutes and defects as shown in Equations (6), (A1) and (A2). A system of equations involving Equations (6), (A1), and (A2) are solved to exactly obtain [*B*], [C], and [V] for each ternary alloy. These concentrations are used to calculate the total transport coefficient matrix ($L_{ij}$) using Equation (5) and the solute diffusion coefficients in Figure 6 using Equation (A5). The concentrations in the dilute regime for Cu(Zr), Cu(Zr,Fe), Cu(Zr,Co) alloys are calculated using a nominal solute concentration of 1at% for each solute species. The resulting concentrations of isolated solute ([*B*], [*C*]), isolated vacancies [V], *B*-*C* pairs [*BC*], and total vacancies ($[\overline{V}]$) are plotted in Figure A1. It is interesting to note that [V*BC*] clusters are not shown here since it overlaps with $[\overline{V}]$ indicating that V-B-C clusters have much higher concentrations than V-B or V-C clusters. To show this effect, the fraction of total vacancies and total solutes in various clusters is calculated and plotted in Figure A2 and A3. The concentration of V-*B*-*C* clusters predominates and therefore the most of vacancies is present in these triplet clusters. Most of the Zr solute (>80%) is present in the B-C clusters due to the high $E_{binding}^{B-C}$. This simplifies this ternary diffusion analyses since *B* and *C* solute will be transported primarily by V-*B*-*C* clusters.

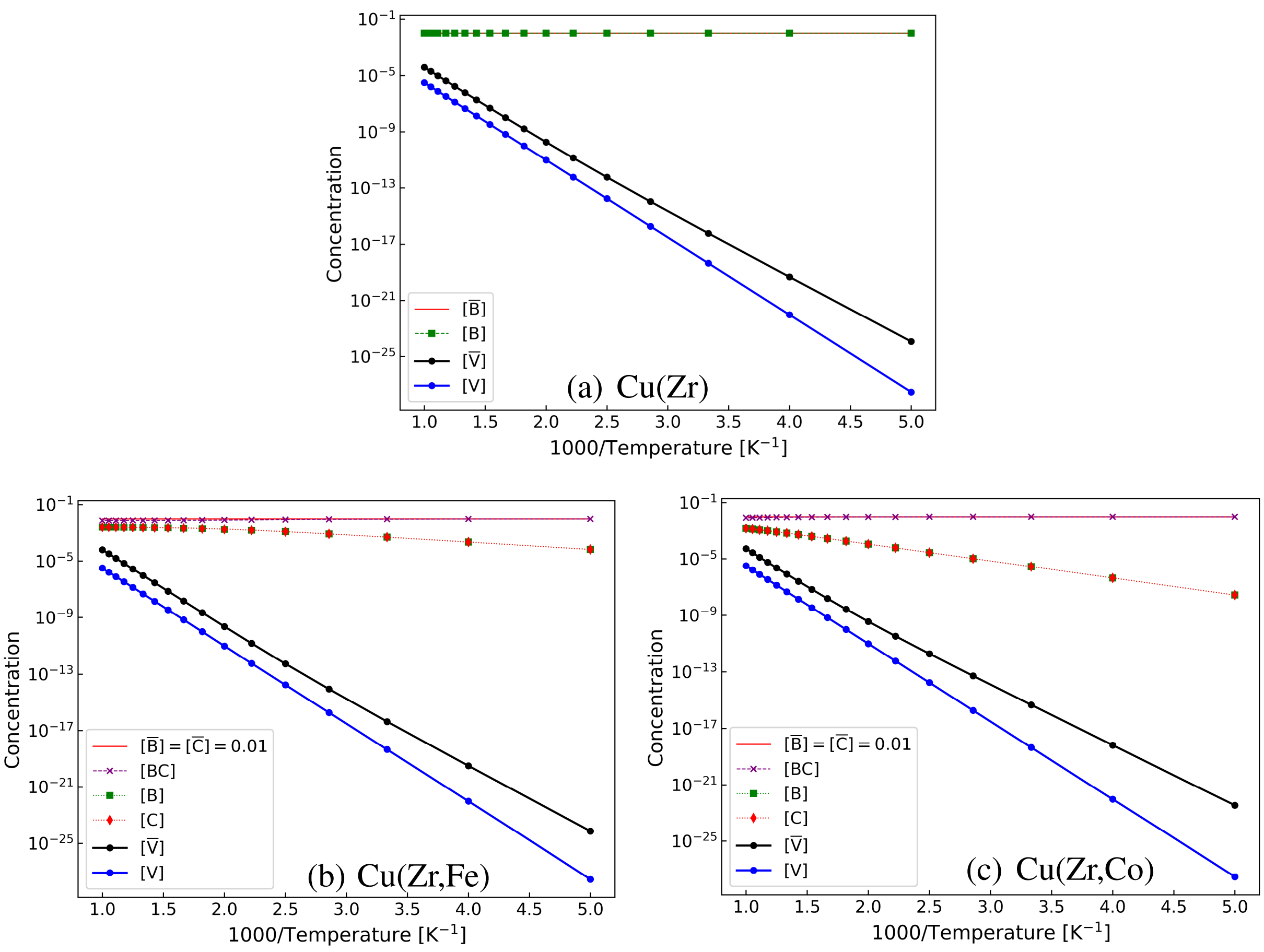

Figure A1: Concentration of solute and clusters in dilute Cu(Zr) alloy in (a), Cu(Zr,Fe) alloy in (b), and Cu(Zr,Co) alloy in (c) under equilibrium conditions for 1at% total nominal concentration of each solute. The total concentration of each solute (including clusters) is notated with an overline (i.e. $[\overline{X}]$). The isolated solute concentrations are notated without an overline (i.e. [X]).

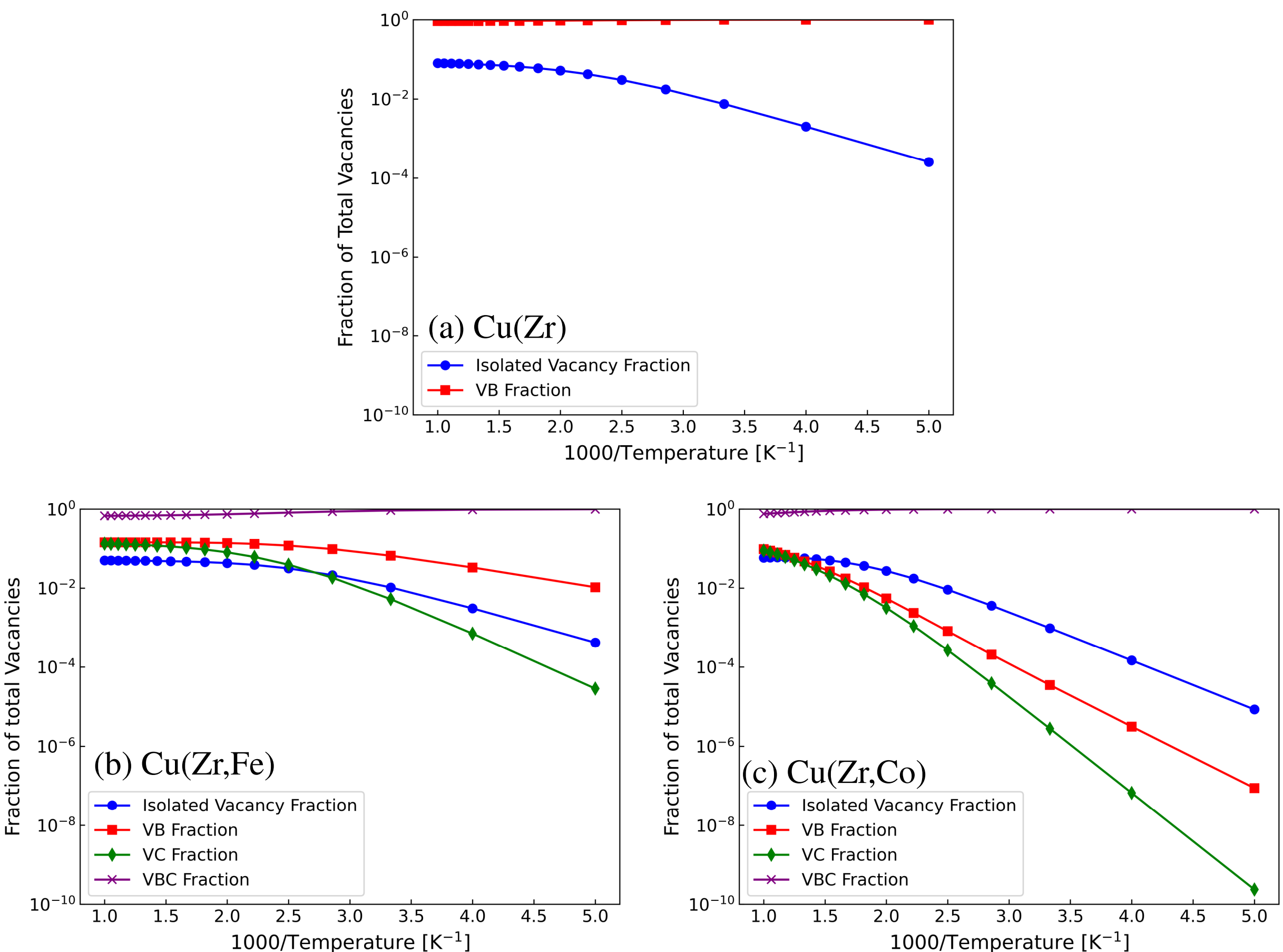

Figure A2: Fraction of vacancies in V-*B*, V-*C*, and V-*B*-*C* clusters in dilute Cu(Zr) alloy in (a), Cu(Zr,Fe) alloy in (b), and Cu(Zr,Co) alloy in (c) under equilibium conditions for 1at% total nominal concentration of each solute

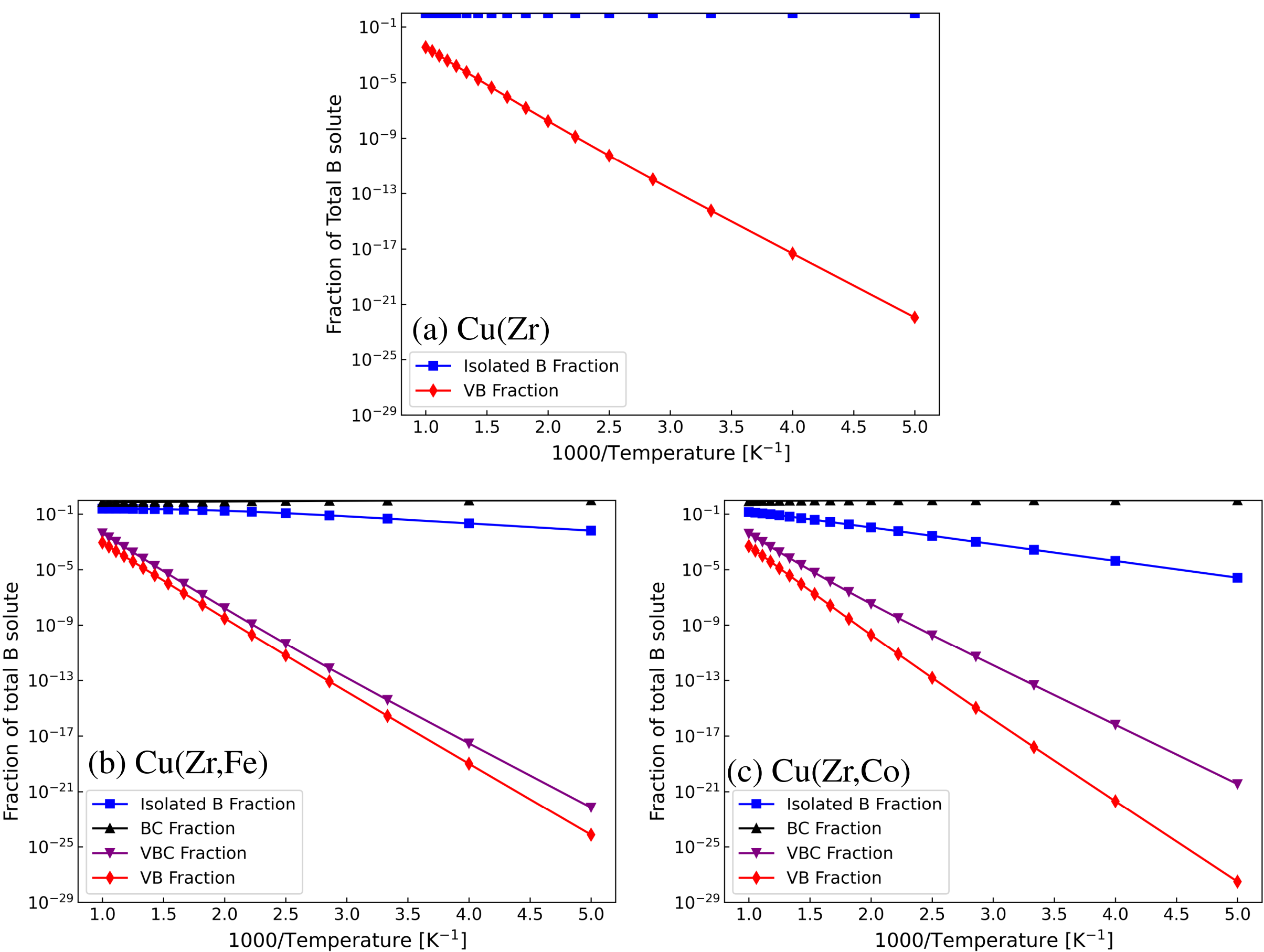

Figure A3: Fraction of nominal Zr solute in V-Zr and V-Zr-*C* clusters in dilute Cu(Zr) alloy in (a), Cu(Zr,Fe) alloy in (b), and Cu(Zr,Co) alloy in (c) under equilibium conditions for 1at% total nominal concentration of each solute

### 3. Solved Concentration of Clusters under irradiation conditions

Similarly to under equilibrium, to obtain the total Onsager matrix for each alloy under non-equilibrium steady state conditions (i.e. under energetic particle irradiation), the concentration of isolated solute, pairs and triplets are calculated by assuming the relative probabilities of clusters follows a Boltzmann equilibrium distribution constrained by the local nominal concentration of solutes and defects in Equations (6), (A1), and (A2). Under energetic particle irradiation, the system of equations is now constrained by the total vacancy concentration determined by Equation (A3). This is numerically solved with Equations (A1) and (A2) to yield the concentrations in Figure (A4) as described in section 1 of the SI using Gekko python package. [49] These concentrations are used to calculate the total transport coefficient matrix of the alloy and thereby the solute diffusion coefficients in Figure 9. The fraction of total vacancies and total solutes in various clusters is calculated and plotted in Figure A5 and A6. Like under equilibrium, the vacancies are mostly in V-*B*-*C* clusters. The Zr solute is mostly in Zr-C clusters in both the Cu(Zr,Fe) and Cu(Zr,Co) alloys. The concentration of total vacancies is largest in the Cu(Zr,Co) alloy consistent with slower diffusion of vacancies in this alloy.

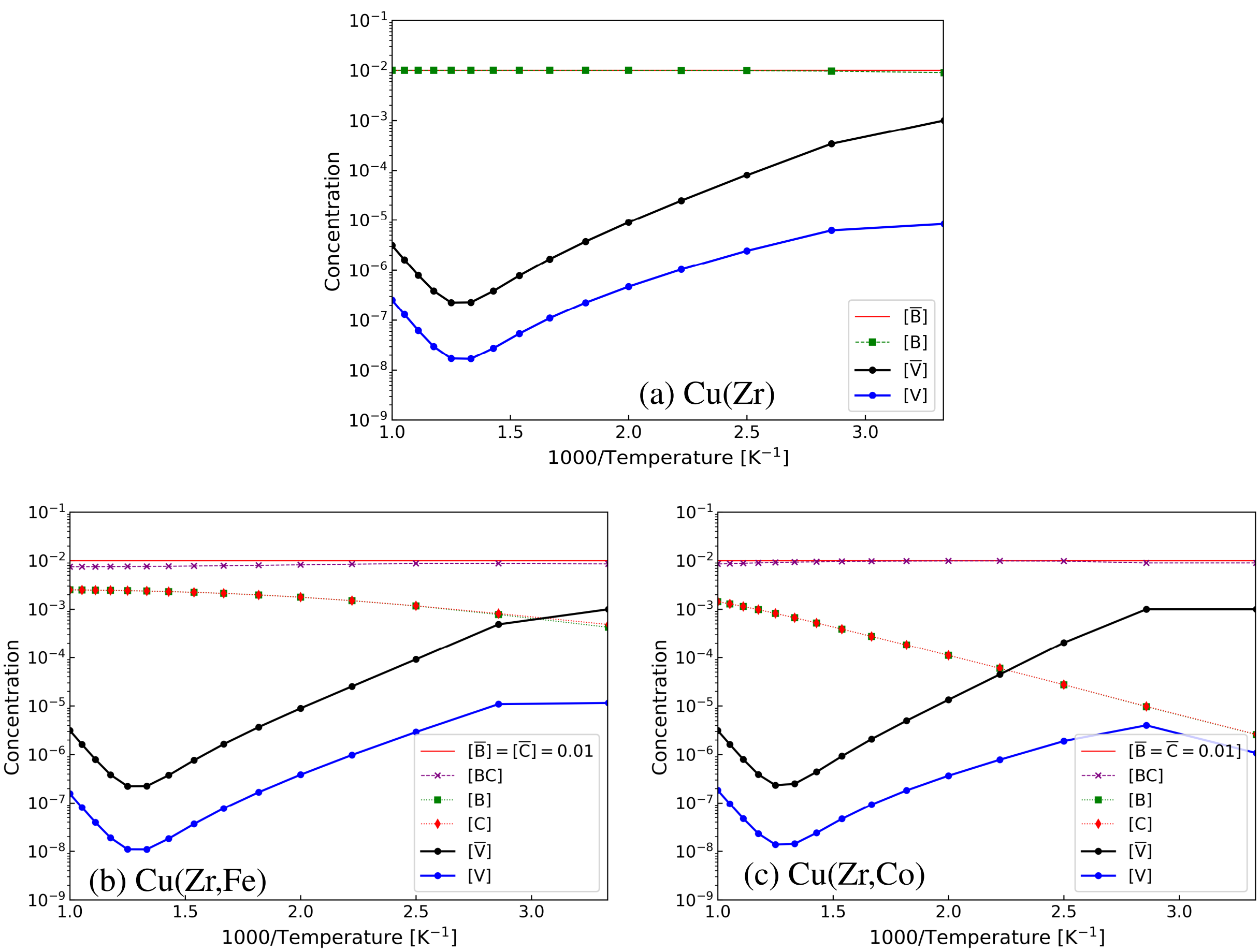

Figure A4**:** Concentration of clusters in dilute Cu(Zr) alloy in (a), Cu(Zr,Fe) alloy in (b), and Cu(Zr,Co) alloy in (c) under nonequilibrium steady state (irradiation) conditions calculated for 1at% total concentration of each solute. The total nominal concentration of each solute is notated with an overline (i.e. $[\overline{X}]$). The isolated solute concentrations are notated without (i.e. [X])

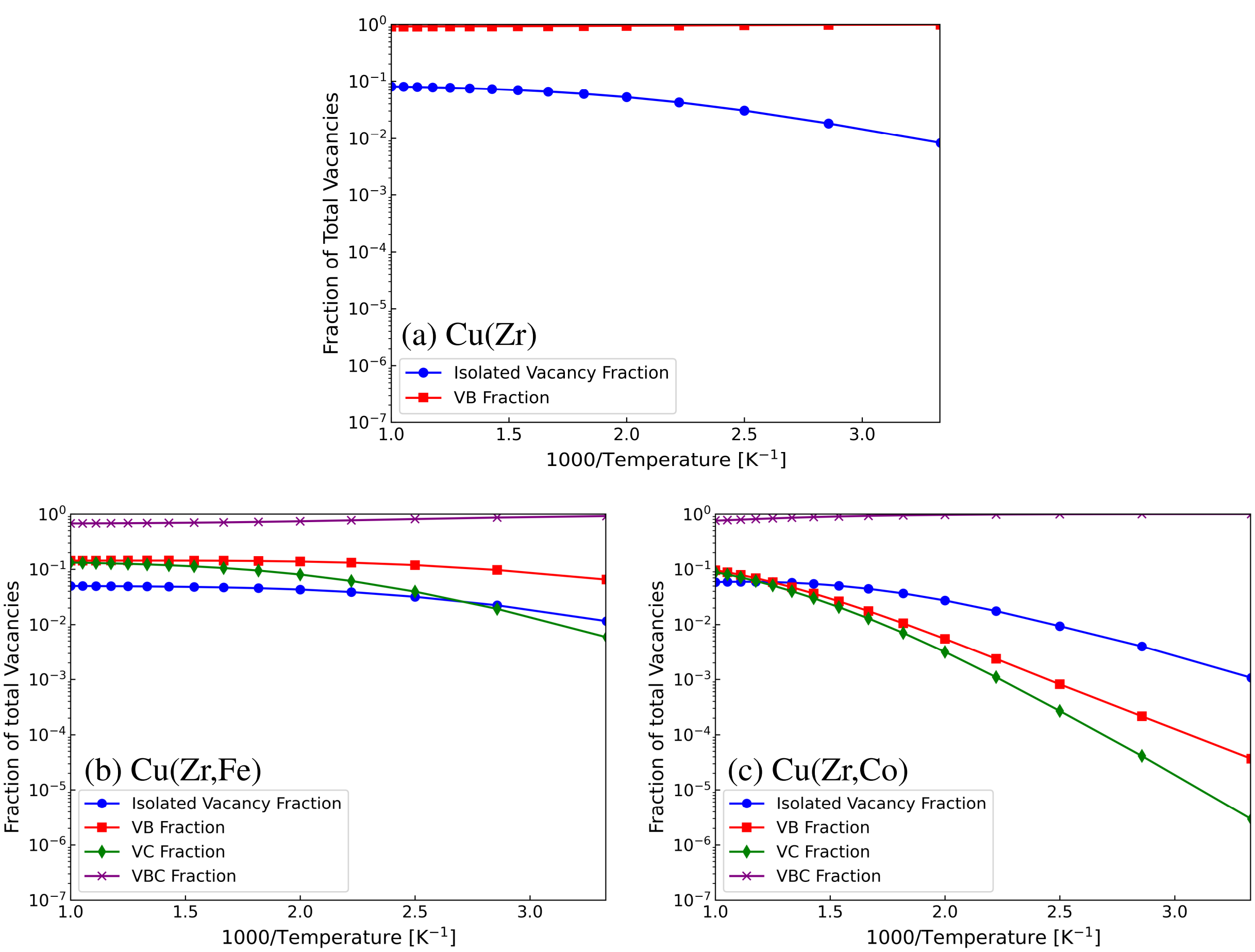

Figure A5: Fraction of vacancies in V-*B*, V-*C*, and V-*B*-*C* clusters in dilute Cu(Zr) alloy in (a), Cu(Zr,Fe) alloy in (b), and Cu(Zr,Co) alloy in (c) under nonequilibrium steady state (irradiation) conditions calculated for 1at% total concentration of each solute.

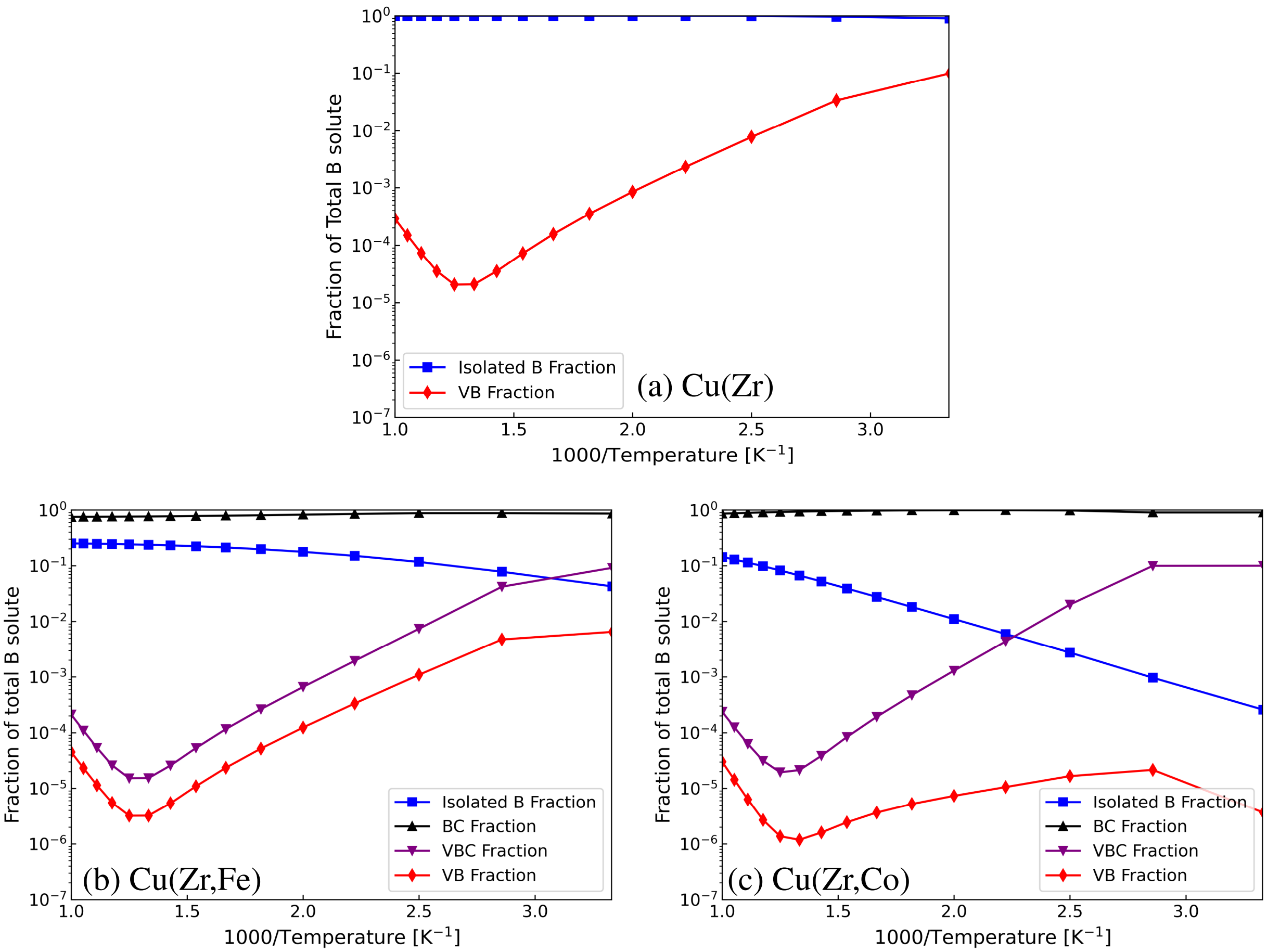

Figure A6: Fraction of Zr solute in V-Zr and V-Zr-C clusters in dilute Cu(Zr) alloy in (a), Cu(Zr,Fe) in (b), and Cu(Zr,Co) in (c) under nonequilibrium steady state (irradiation) conditions calculated for 1at% total concentration of each solute.

The normalized total vacancy concentration is compared between Cu(Zr), Cu(Zr,Fe) and Cu(Zr,Co) in Figure A7 which shows that total vacancy concentration is increased slightly at lower temperatures between the pure Cu and Cu(Zr,Co) alloy. The increased vacancy concentration at higher temperatures (~750K) is due to increased vacancy concentration at equilibrium due to the large vacancy binding energy, see Eqs. 7 and A1. This term is additionally increased for Cu(Zr,Co) compared to Cu(Zr) due to the larger effective vacancy binding energy which increases the total concentration of vacancies. The increase in vacancy concentration at lower temperatures (~400K) is due to the reduced vacancy diffusion in the respective systems. For Cu(Zr,Fe) and Cu(Zr,Co), the vacancies are effectively trapped in V-Zr-*C* clusters slowing their mobility and thus increasing the overall concentration of vacancies. This effect is more pronounced in Cu(Zr,Co) compared to Cu(Zr,Fe) due to much lower diffusivity of vacancies in V-Zr-Co clusters compared to V-Zr-Fe clusters as mentioned in the Results.

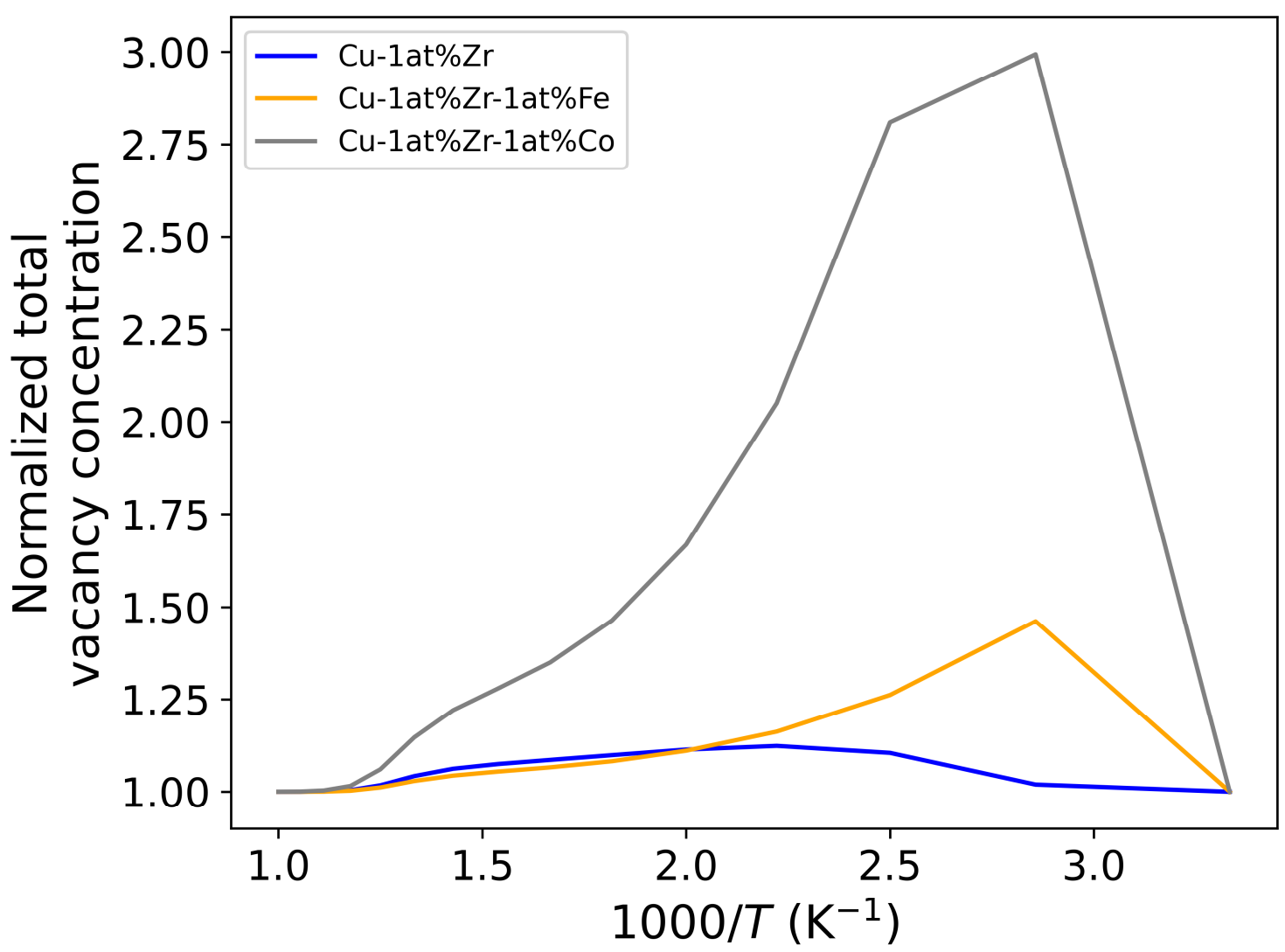

Figure A7: Total vacancy concentration for Cu(Zr), Cu(Zr,Fe), and Cu(Zr,Co) alloys normalized by vacancy concentration in pure Cu, modeled under irradiation conditions. Vacancy concentration is increased in ternary alloys compared to binary due to the reduced vacancy diffusion coefficient.

### 4. Fe and Co Solute Diffusion in dilute Cu(Zr,Fe) and Cu(Zr,Co) alloys

Solute diffusion coefficients of the *C* solute in Cu(Zr,Fe) and Cu(Zr,Co) are shown in Figure A8. The presence of V-Zr-*C* clusters seems to increase the C solute diffusivity in these cases since the Fe and Co are now transported in these 3-body clusters. For Cu(Zr,Fe), the activation energy for solute diffusion is decreased by ~0.2eV from the Cu(Zr) binary to 1.686 eV. For Cu(Zr,Co), a similar decrease is observed, but since Co is more immobile than Fe, the solute diffusion is still slower than self-diffusion of Cu.

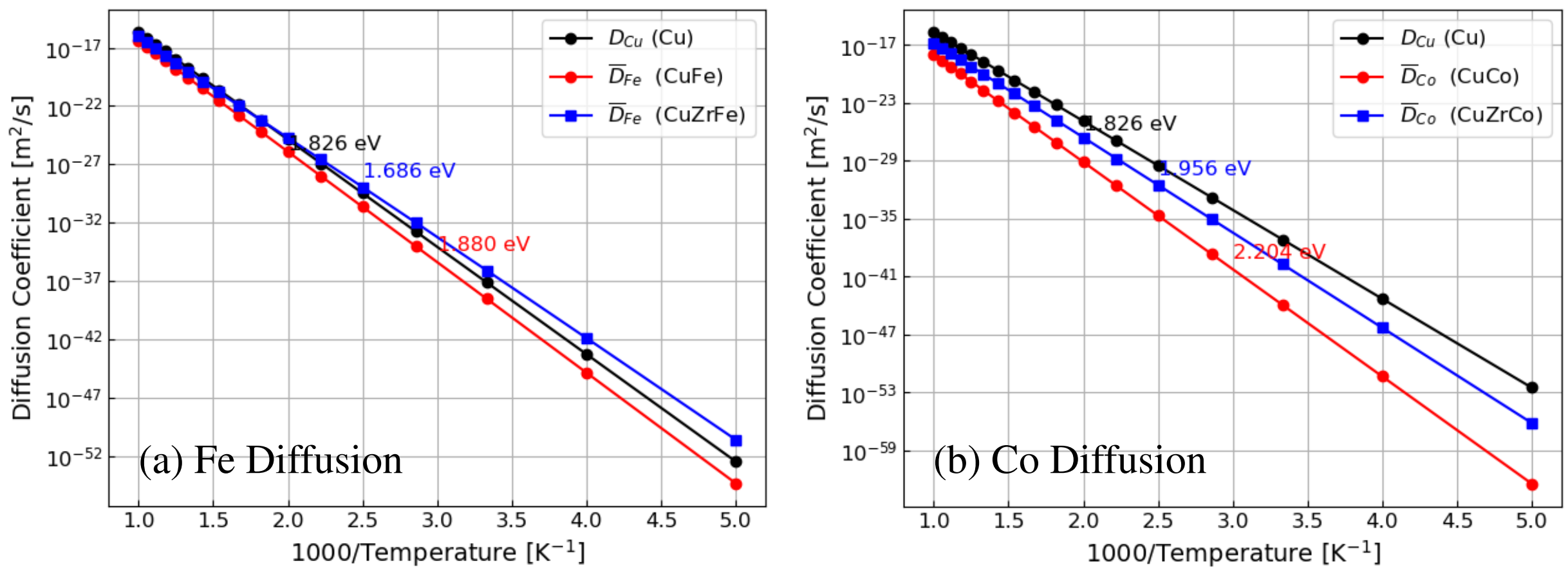

Figure A8: Equilibrium C-type solute diffusion coefficients for dilute Cu(Fe) vs Cu(Zr,Fe) alloys in (a), and dilute Cu(Co) vs Cu(Zr,Co) alloys in (b) calculated at 1at% for each solute concentration

### 5. Activation energy corrections for comparison to literature

To facilitate comparison with the diffusion data reported by H. Wu et al. [28], the activation energies for solute diffusion are corrected according to the experimental value for matrix self-diffusion following their methodology: The self-diffusion activation energy is first corrected to its experimental value (2.080 eV for pure Cu), and the resulting correction offset is then uniformly applied to the activation energies of all other solutes. The activation energy correction offset for the data presented here is 0.254 eV. The uncorrected and corrected activation energies are reported in Table III.

Table III: Uncorrected and corrected activation energies for solute diffusion in dilute binary Cu(Zr) and ternary Cu(Zr,Fe) and Cu(Zr,Co) alloys with 1 at% nominal concentration for each solute

| Solute | Alloy | Uncorrected Q [eV] | Corrected Q [eV] |
|---|---|---|---|
| Cu | Cu | 1.826 | 2.080 |
| Zr | Cu(Zr) | 1.574 | 1.828 |
| Zr | Cu(Zr,Fe) | 1.706 | 1.960 |
| Zr | Cu(Zr,Co) | 1.821 | 2.075 |
| Fe | Cu(Fe) | 1.880 | 2.134 |
| Fe | Cu(Zr,Fe) | 1.686 | 1.940 |
| Co | Cu(Co) | 2.204 | 2.458 |
| Co | Cu(Zr,Co) | 1.956 | 2.210 |

## APPENDIX B: ADDITIONAL ANALYSIS OF DFT RESULTS

### 1. Dilute binary Cu(Zr), Cu(Fe), and Cu(Co) migration energies computed using DFT

The migration energies for vacancies near Zr, Fe, and Co are computed using Quantum Espresso [34, 35] with PSLibrary PAW Pseudopotentials [38] in Table IV compared to H. Wu et al. which are computed using VASP's PAW pseudopotentials. [28] The migration energies for Cu(Zr) agree well with this reference. However, for Cu(Fe) and Cu(Co), there is a disagreement especially with $E_2^m$ (~0.15 eV difference). This is due to the magnetic moment of the solute after energy minimization. A local minimum of the magnetic moment is observed at 0 μB, yielding approximately the same migration energies calculated by H. Wu et al. However, nonzero magnetic moments for the Fe/Co solute (2.7 μB for Fe and 1.5 μB for Co) lead to a lower energy state. This observation matches well with another DFT calculation of Fe solutes in Cu calculated by Eisenbach [49]. The reduction in $E_2^m$ lowers the activation energy for diffusion of the Fe solute. It is much closer to the self-diffusion coefficient of Cu.

The unique behavior of vacancy solute drag in the Cu(Zr) alloy stems from the interplay of specific vacancy migration energies near the Zr solute. For Cu(Zr), the five frequencies for vacancy migration are listed in Table IV. There is an especially fast reassociation ($E_4^m$=0.48eV) of vacancies from nearby sites (2NN, 3NN, 4NN) to 1NN sites to the Zr atom. Once a Zr atom is 1NN with a vacancy, there is rapid switching with the vacancy due to the exceptionally low V-Zr swap energy ($E_2^m$=0.26eV). Then, the vacancy will dissociate ($E_3^m$=0.66eV) more frequently than rotating around the solute ($E_1^m$=1.20eV). Once dissociating to 2NN, 3NN, or 4NN position to the Zr atom, the vacancy can reassociate to the Zr atom, resulting in an effective rotational motion of the vacancy around the Zr atom if this is a different 1NN position than initially. This occurs preferentially to the vacancy diffusing away from the Zr atom due to the lower migration barrier associated with this jump. Altogether, these kinetic correlations cause the Zr atom to diffuse with the vacancy to sinks, causing solute drag in the Cu(Zr) alloy [18,19,22].

Table IV: Vacancy migration energies in dilute binary Cu(Zr), Cu(Fe), and Cu(Co) alloys computed using DFT and compared to Ref. [27].

| | | **Zr** | | **Fe** | | **Co** | |
|---|---|---|---|---|---|---|---|
| (units=eV) | | This work | Ref. [28] | This work | Ref. [28] | This work | Ref. [28] |
| $E_0^m$ | | 0.734 | 0.717 | 0.734 | 0.717 | 0.734 | 0.717 |
| $E_1^m$ | | 1.201 | 1.193 | 0.765 | 0.670 | 0.710 | 0.664 |
| $E_2^m$ | | 0.266 | 0.262 | 0.781 | 1.031 | 0.880 | 1.040 |
| $E_3^m$ | 1NN↔2NN | 0.742 | - | 0.708 | - | 0.704 | - |
| $E_4^m$ | | 0.454 | - | 0.749 | - | 0.791 | - |
| $E_3^m$ | 1NN↔3NN | 0.619 | - | 0.716 | - | 0.724 | - |
| $E_4^m$ | | 0.440 | - | 0.773 | - | 0.813 | - |
| $E_3^m$ | 1NN↔4NN | 0.661 | 0.647 | 0.726 | 0.729 | 0.745 | 0.731 |
| $E_4^m$ | | 0.482 | 0.467 | 0.752 | 0.814 | 0.795 | 0.802 |

**2. Magnetic moments of solutes Fe and Co during Solute-Solute binding and Solute-Vacancy Binding**

The Fe and Co solutes changed magnetic moments during interactions with vacancies and the Zr solute. This magnetic moment is obtained by starting the DFT calculation with several initial magnetic moments, and then the structure with the lowest energy is chosen as the ground state. These magnetic moments are documented in Table V. When the Fe or Co solute binds with a vacancy in bulk Cu, the magnetic moment is increased. When the Fe or Co solute atoms bind with the Zr solute, the magnetic moment is lowered. For Co, the magnetic moment is almost completely nullified.

Cu atoms are non-magnetic, so there is only one atom (the Fe solute) which may have a non-zero magnetic moment. Initial magnetic moments for the Fe atom were tested (0, 1, 2, and 3 μB) in both the isolated and bound state with 1NN Zr in the copper matrix. The Zr-Fe binding energy in copper is 0.210 eV when the Fe converges to a magnetic moment of 2.2561 μB. The Zr-Fe binding energy is increased to 0.612 eV when the Fe atom converges to a magnetic moment of 0.0005 μB, but this configuration is not energetically favorable (by ~0.2 eV) compared to the configuration where Fe atom has a magnetic moment of 2.2561 μB with a 1NN Zr solute. Therefore, the binding energy is 0.210 eV for 1NN Zr-Fe pairs in the copper matrix with the magnetic moments listed in Table V for the isolated Fe solute and 1NN Fe-Zr pairs.

Table V: Magnetic moment of Fe and Co solute during pair interactions in the copper matrix

| | Fe solute Magnetic moment (μB) | | Co solute Magnetic moment (μB) | |
|---|---|---|---|---|
| Configuration of Magnetic Solute | single Fe solute | 2.6669 | single Co solute | 1.4091 |
| | V-Fe 1NN pair | 2.7713 | V-Co 1NN pair | 1.5168 |
| | Fe-Fe 1NN pair | 2.5923 | Co-Co 1NN pair | 1.4521 |
| | Fe-Zr 1NN pair | 2.2561 | Co-Zr 1NN pair | 0.0008 |

## APPENDIX C: EFFECTIVE BINDING ENERGY AND JUMP FREQUENCIES FOR COMPARING BINARY AND TERNARY ALLOYS

**1. Effective binding energy of vacancies and triplet corrections for ternary clusters:**

These configurations of triplet V-Zr-*C* clusters in Figure C1 have the highest binding energies and are identified using KineCluE from the initial pairwise binding energies. All the configurations have 1NN between V-Zr and Zr-*C* pairs. The distance between the vacancy and *C* solute are 1NN in the (111) triangle, 2NN in the (100) triangle, 3NN in the bent chain configuration, and 4NN in the linear chain configuration.

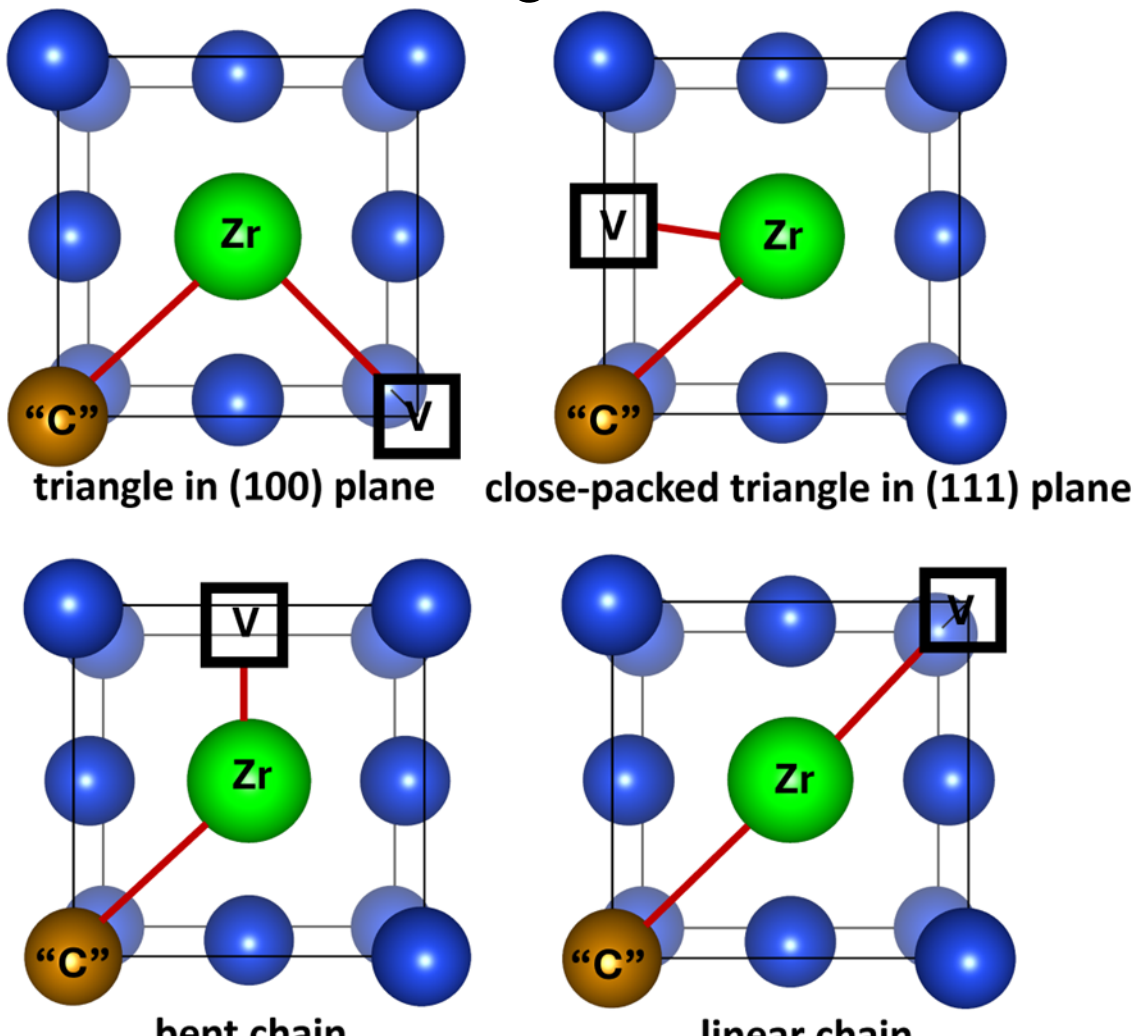

Figure C1: Configurations of triplet V-Zr-*C* clusters in FCC Cu used for binding energy calculations by DFT

The Fe and Co solutes adopted various magnetic moments during the interactions with vacancies and the Zr solute in these triplet configurations as listed in Table VI. For both the V-Zr-Fe and V-Zr-Co clusters, the magnetic solute had lower magnetic moments correlated with higher binding energies. This indicates that these configurations enhance the vacancy attraction to the Zr-C pair such as with the (111) triangle configurations. The only exception is with the Co solute in the (100) triangle which had the lowest magnetic moment but still had a lower triplet binding energy than expected using the sum of pairwise interactions up to 5NN.

Table VI**:** Magnetic moment of Fe and Co solute in the triplet V-Zr-*C* clusters in FCC Cu

| | Fe solute Magnetic moment (μB) | | Co solute Magnetic moment (μB) | |
|---|---|---|---|---|
| Configuration | (100) triangle | 2.2084 | (100) triangle | 0.0002 |
| | (111) triangle | 1.9864 | (111) triangle | 0.0026 |
| | linear bent | 2.3906 | linear bent | 0.0782 |
| | linear | 2.4487 | linear | 0.0435 |

The effective binding energy of vacancies to Zr-C clusters is calculated by considering the previous triplet configurations. KineCluE provides the geometrical multiplicities for these configurations given in Table VII, so a thermodynamic average of the triplet binding energies is calculated using Equation (C1).

$$E_{binding}^{V-B-C} = \log\left(\frac{1}{\sum_{\rho\in\rho_{V-B-C}} g_{V-B-C}} z_{V-B-C}\right) \tag{C1}$$

where $g$ is the geometric multiplicity of the configuration ρ, and z is the partition function for the cluster.

Then, an effective binding energy of the vacancy to the Zr-*C* pair can be defined by Equation (C2).

$$E_{eff.\ binding}^{V-BC} = E_{binding}^{V-B-C} - E_{binding}^{B-C} \tag{C2}$$

where $E_{eff.\ binding}^{V-BC}$ is the effective binding energy of the vacancy to the Zr-*C* pair, $E_{binding}^{V-B-C}$ is the binding energy of the V-*B*-*C* clusters, and $E_{binding}^{B-C}$ is the binding energy of *B*-*C pairs.*

This gives us a comparison of the vacancy attraction to solutes between the pair and triplet clusters. The result shows that for Zr-Fe pairs, the vacancy attraction is reduced by 18-20 meV in the 500-650K range. For the Zr-Co pairs, the vacancy attraction is increased by 2-9 meV in the temperature range. Additionally, it is evident that the highest binding energy configuration associated with the (111) triangle has the highest weight in this thermodynamic average of triplet binding energies due to the exponential of the binding energies in the partition function equation and the geometric multiplicity of this configuration (see Equation 6).

Table VII**:** Geometric multiplicities of triplet V-Zr-*C* configurations identified using KineCluE

| | | Geometric Multiplicity, $g_\rho$ |
|---|---|---|
| Configuration | (100) triangle | 24 |
| | (111) triangle | 48 |
| | linear bent | 12 |
| | linear | 48 |

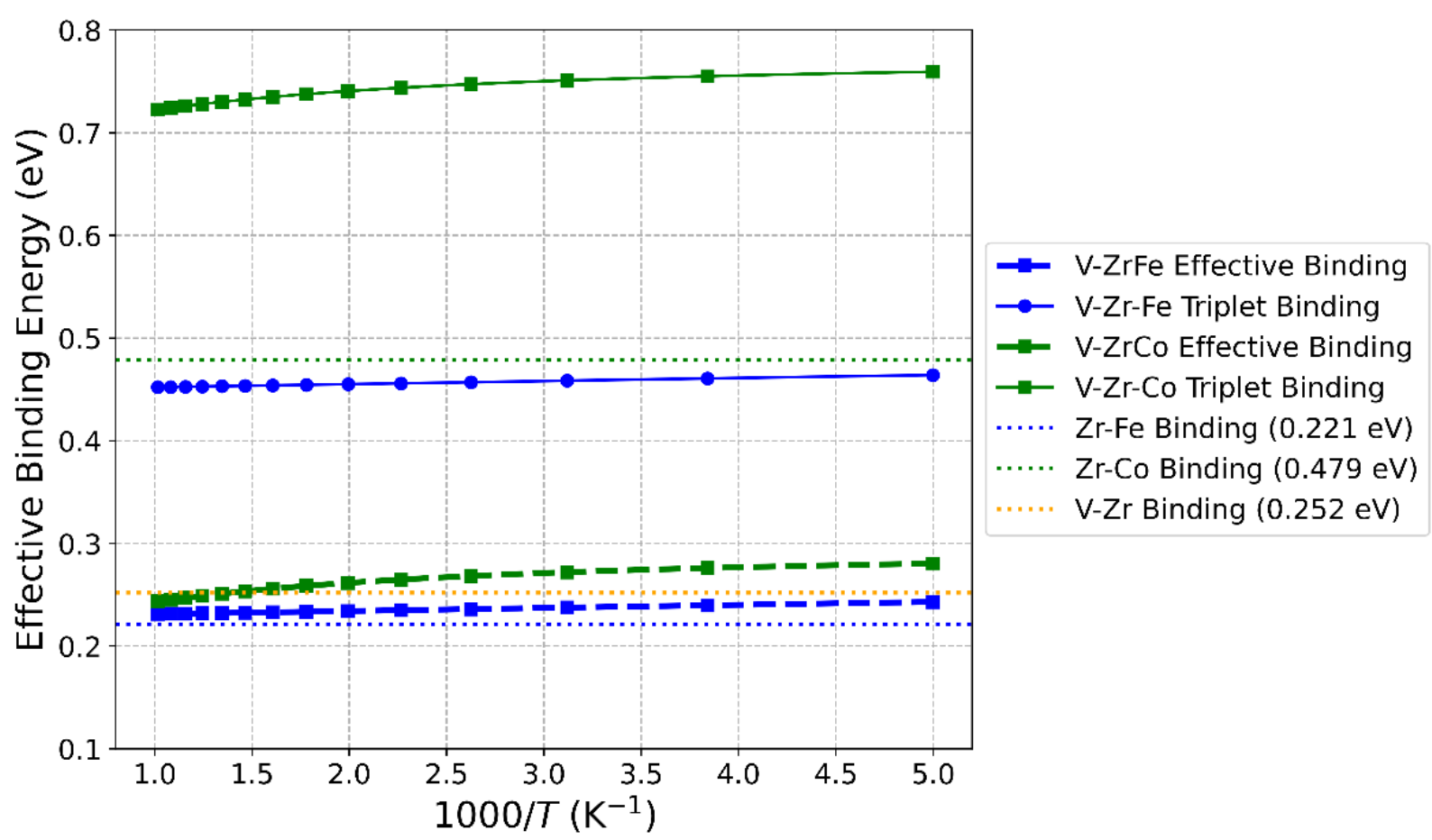

Figure C2: Effective binding energy of vacancy to Zr-*C* pairs calculated from a thermodynamic average of V-Zr-*C* cluster binding energies from DFT.

## 2. Effective jump frequencies for the ternary alloy within the five-frequency framework

The analytical expressions for Onsager transport coefficients in the five-frequency framework is expressed using a polynomial function of normalized jump frequencies ($W_i = w_i p_i / w_0 p_0$) [17,19]. After substituting the jump frequencies for Zr into the analytical expressions for the transport coefficients, it becomes evident that $W_3$ is the most important term for establishing the magnitude of $L_{ZrV}/L_{VV}$ due to the fifth-order polynomial of $W_3$. For Cu(Zr), $W_2$ almost entirely cancels out when calculating $L_{ZrV}/L_{VV}$ due to the magnitude of $W_3 P_3$ compared to $W_1 P_4 + W_3 P_5$ term in the $L_{VV}$ equation. Additionally, $W_1 P_0$ becomes insignificant compared to $W_3 P_1$ in $L_{VB}$ equation due to the high barrier for $E_1^m$.

Due to the reduction in effective binding energy to vacancies for the V-Zr-Fe cluster as noted in Figure C2, the $L_{BV}/L_{VV}$ heatmaps are recalculated from Figure C3 with a reduced binding energy of 0.162 eV showing the location of the Cu(Zr,Fe) ternary alloy in Figure C3. Combined with Figure 12, we can deduce that about half the reduction in solute drag (0.05) from Cu(Zr) to Cu(Zr,Fe) is due to the reduction in effective binding energy to vacancies which is undesirable for the application of vacancy trapping under irradiation.

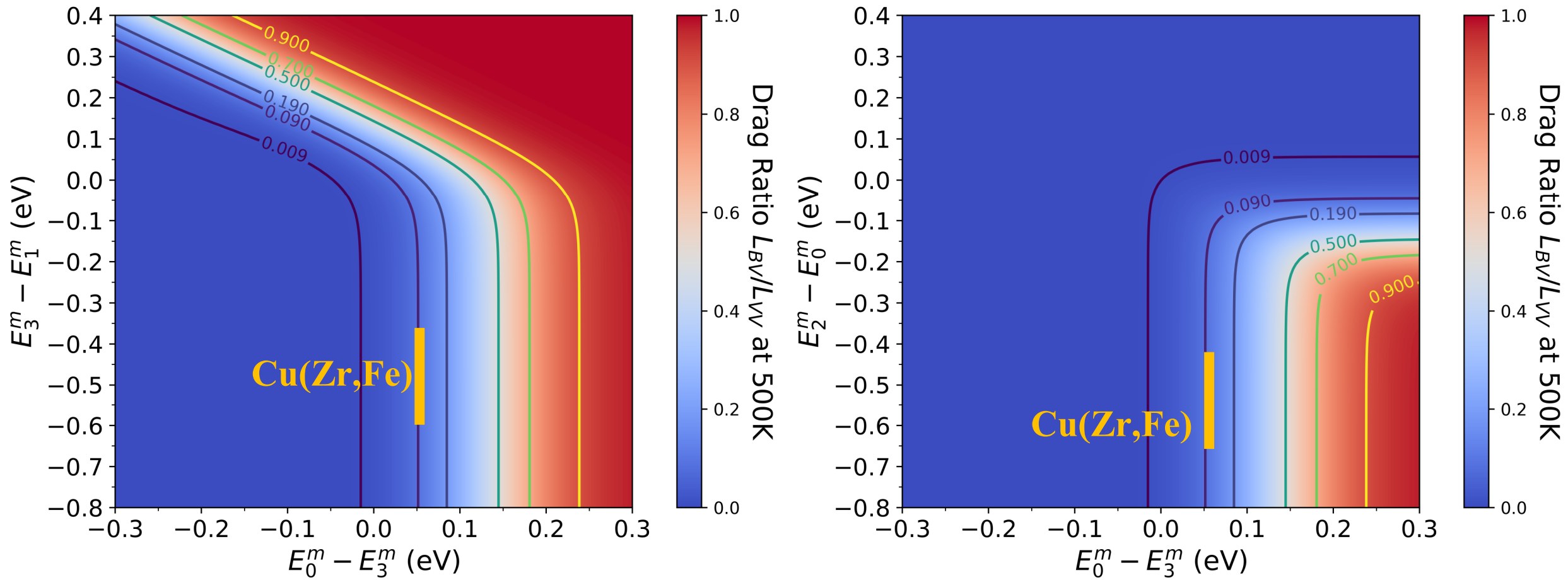

Figure C3: Solute drag maps ($L_{BV}/L_{VV}$) showing location of ternary Cu(Zr,Fe) at 500K calculated with a reduced $E_{binding}^{B-V} = 0.163\ eV$ representative of the effective vacancy binding energy to Zr-Fe pairs.

Figure C4 shows that one of the $\omega_1$-type migration energy for the V-Zr-Fe cluster is raised to 1.257 eV compared to the 1.201 eV of the V-Zr cluster. This indicates that our assumptions are valid that $\omega_1$ is not significantly increased from the Cu(Zr) to Cu(Zr,Fe) and Cu(Zr,Co) alloys in Figure 12 and C3. A similar calculation for V-Zr-Co shows that the migration energy barrier has increased to 1.292 eV in that case.

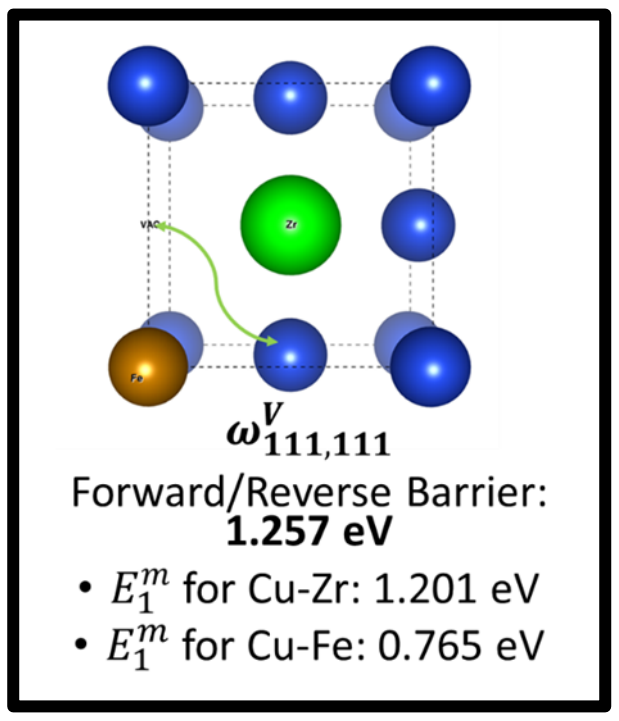

Figure C4: $w_1$-type migration energy for V-Zr-Fe cluster compared to the jump frequencies for the binary Cu(Zr) and Cu(Fe) alloys

The solute diffusion coefficient, taken from Philibert et al. [54], is shown in Equation (S11) using the five-jump frequency model for FCC crystals. This analytically matches the solute diffusion coefficient calculated using KineCluE calculations with Equation (S5).

$$D_B^A = a^2 \overline{N_v} \frac{w_4}{w_3} \cdot \frac{w_1 + (7/2)Fw_3}{w_1 + w_2 + (7/2)Fw_3} \cdot w_2 \qquad (C3)$$

In this equation, F is a polynomial function of jump frequencies. In the case of solutes with large vacancy binding energy (>0.05 eV), this function simplifies to 7F=2. Furthermore, solutes with large vacancy-binding energy in the copper matrix also have high $w_2$ jump frequencies. Since $w_3 \approx w_0$ for all solutes based on previous calculations, $w_2$ becomes the dominant denominator term which mostly cancels out with $w_2$ in the numerator in Equation (C3). Therefore, $D_B^A \approx a^2\overline{N_v}\frac{w_4}{w_3}w_0 = a^2\overline{N_v}w_0E_B^{V-B}$ when the vacancy-binding energy is large. The activation energy for solute diffusion therefore becomes $Q_B^A \approx Q_{SD} - E_B^{V-B}$, where $Q_{SD}$ is the activation energy for the self-diffusion of copper. This equation is the origin of the inverse relationship with a slope of -1 between vacancy-binding energy and solute activation energy for diffusion for strong vacancy binding solutes. Additionally, this linear fitting goes through the self-diffusion point for Cu due to the prefactor term. Remarkably, for large enough solute binding energy solutes, this linear relationship is observed in Al, Ni, and Fe host matrices using the data from Ref. [27]. It should be noted, however, that the threshold in binding energy varies for this relationship to hold in each matrix.

## ACKNOWLEDGMENTS:

This research is supported by the NSF under Grant No. DMR- 2105118. We gratefully acknowledge stimulating discussions with Dr. Soumyajit Jana. V.V. was supported by the NSF under Grant No. DGE-192275. This work made use of the Illinois Campus Cluster, a computing resource that is operated by the Illinois Campus Cluster Program (ICCP) in conjunction with the National Center for Supercomputing Applications (NCSA) and which is supported by funds from the University of Illinois at Urbana-Champaign. Additionally, this work used the DELTA cluster at the NCSA through allocation MAT230076 from the Advanced Cyberinfrastructure Coordination Ecosystem: Services & Support (ACCESS) program, which is supported by U.S. National Science Foundation grants #2138259, #2138286, #2138307, #2137603, and #2138296.

# DFT-informed Design of Radiation-Resistant Dilute Ternary Cu Alloys

Vaibhav Vasudevan[1], Thomas Schuler[2], Pascal Bellon[1], Robert Averback[1]
[1] Materials Research Laboratory, University of Illinois Urbana-Champaign 104 South Goodwin Ave. MC-230 Urbana, IL 61801
[2] Université Paris-Saclay, CEA, Service de recherche en Corrosion et Comportement des Matériaux, SRMP, 91191 Gif-sur-Yvette, France

## SUPPLEMENTAL MATERIAL:

### 1. Approximation of thermodynamic range in Cu(Zr), Cu(Zr,Fe) and Cu(Zr,Co) alloys

The error associated with the 1NN thermodynamic range is tabulated in Table S1 and compared to 5NN thermodynamic range on the triplet binding energies and on the solute drag ratio calculated using KineCluE. The triplet binding energy corrections for (100) and (111) compact triangles configurations with 1NN V-Zr and Zr-*C* are quite small compared to linear bend and linear configurations. From the previous analysis on the effective binding energy in Figure S2, the highest binding energy configuration has the largest contribution to this effective binding energy due to the exponential in the partition function. Therefore, the higher triplet correction associated with linear and linear bent configurations may have a lesser impact on the effective binding energy compared to the (100) and (111) triangle configurations.

Table S1**:** Binding Energy Error associated with 1NN thermodynamic range calculations compared to 5NN

| | **V-Zr-Fe Clusters** | | **V-Zr-Co Clusters** | |
|---|---|---|---|---|
| | 1NN Pairwise Binding Energy | Error in eV (Percent) | 1NN Pairwise Binding Energy | Error in eV (Percent) |
| (100) triangle | 0.473 eV | 0.015 eV (3.1%) | 0.731 eV | 0.002 eV (0.2%) |
| (111) triangle | 0.431 eV | 0.022 eV (5.1%) | 0.661 eV | 0.115 eV (17.4%) |
| linear bent | 0.473 eV | 0.040 eV (8.4%) | 0.731 eV | 0.122 eV (16.7%) |
| linear | 0.473 eV | 0.050 eV (10.5%) | 0.731 eV | 0.160 eV (21.9%) |

The effect of this error on the KineCluE calculations was considered in Figure S1 assuming binding energies of three-body clusters are a sum of pairwise interactions for both binary Cu(Zr) and ternary Cu(Zr,Fe) systems using 1NN and 5NN thermodynamic range with KRA method for jump frequencies. This analysis reveals that extending the thermodynamic range to 5NN results in a greater solute drag effect compared to the 1NN range in both systems. Nevertheless, to establish an initial understanding of the kinetic origins of solute drag within the ternary systems, this paper utilizes the 1NN thermodynamic range to understand the effects of the *B*-*C* binding energy. This approach serves to limit the exploration of numerous lower binding energy configurations which are unlikely to affect the triplet diffusion mechanisms, thereby simplifying the initial analyses.

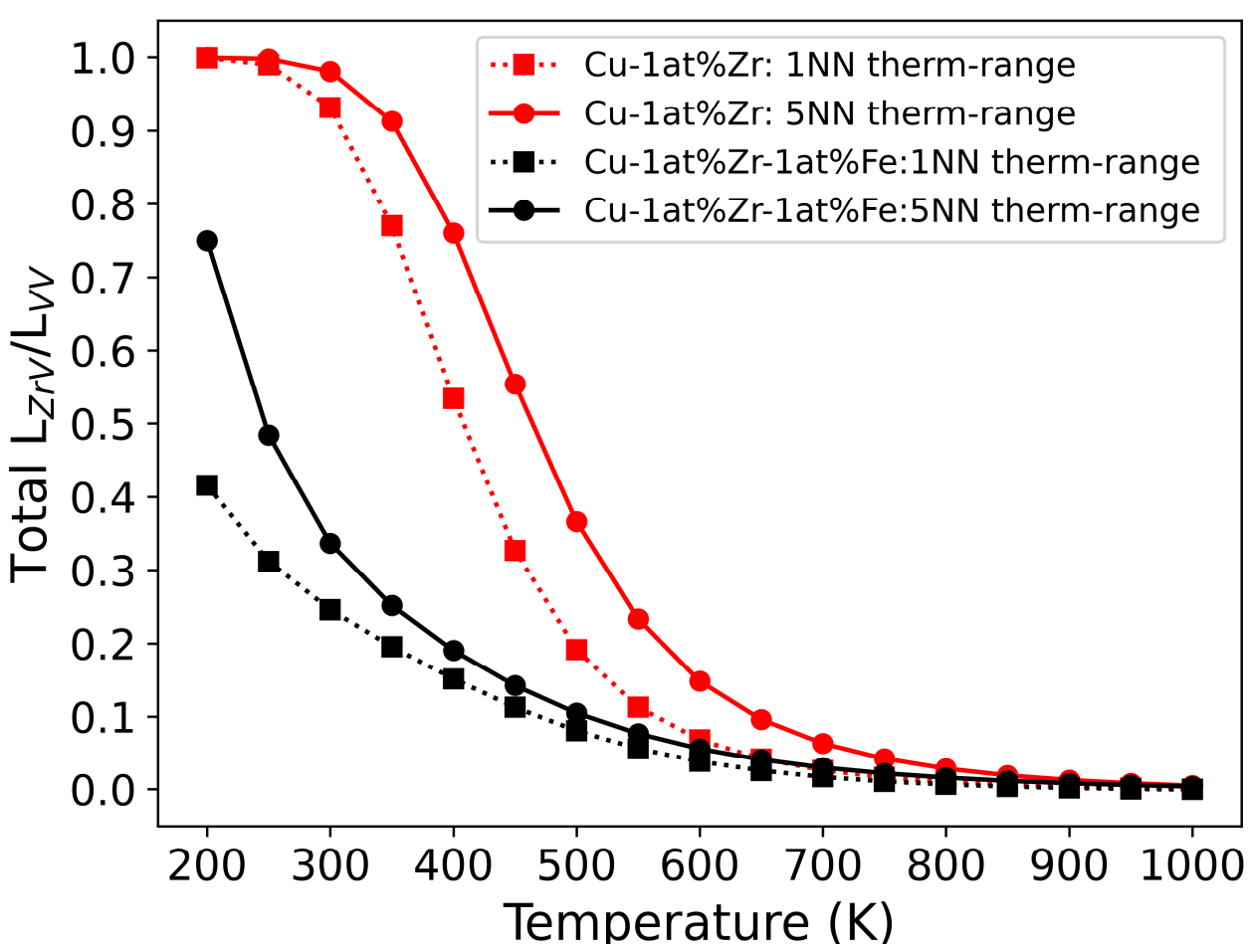

Figure S1: Effect of 1NN and 5NN thermodynamic range on vacancy solute drag ratio $L_{VZr}/L_{VV}$ in Cu(Zr) and Cu(Zr,Fe) alloys with a nominal solute concentration of 1 at%

## 2. Kinetic range analysis for KineCluE calculations

To analyze the effect of the kinetic range approximation in the KineCluE calculations, a convergence study was performed on the V-Zr-Fe cluster from 1.0a to 5.5a as shown in Figure S2. As the kinetic range increases, the computational demand grows exponentially as the number of configurations and jump frequencies explored in the larger volume grows exponentially as shown in Figure S3. The largest calculation with a kinetic range of $5.5a_0$ took nearly 244 hours and 45 GB in memory per calculation (using 1 core of an AMD EPYC 7713 node) compared to only 5.56 hours and 5 GB in memory per calculation for the $4.0a_0$ kinetic range, while still providing meaningful results. Based on this, a kinetic range of $4.0a_0$ was chosen for the pair and triplet calculations to obtain comparable and converged results.

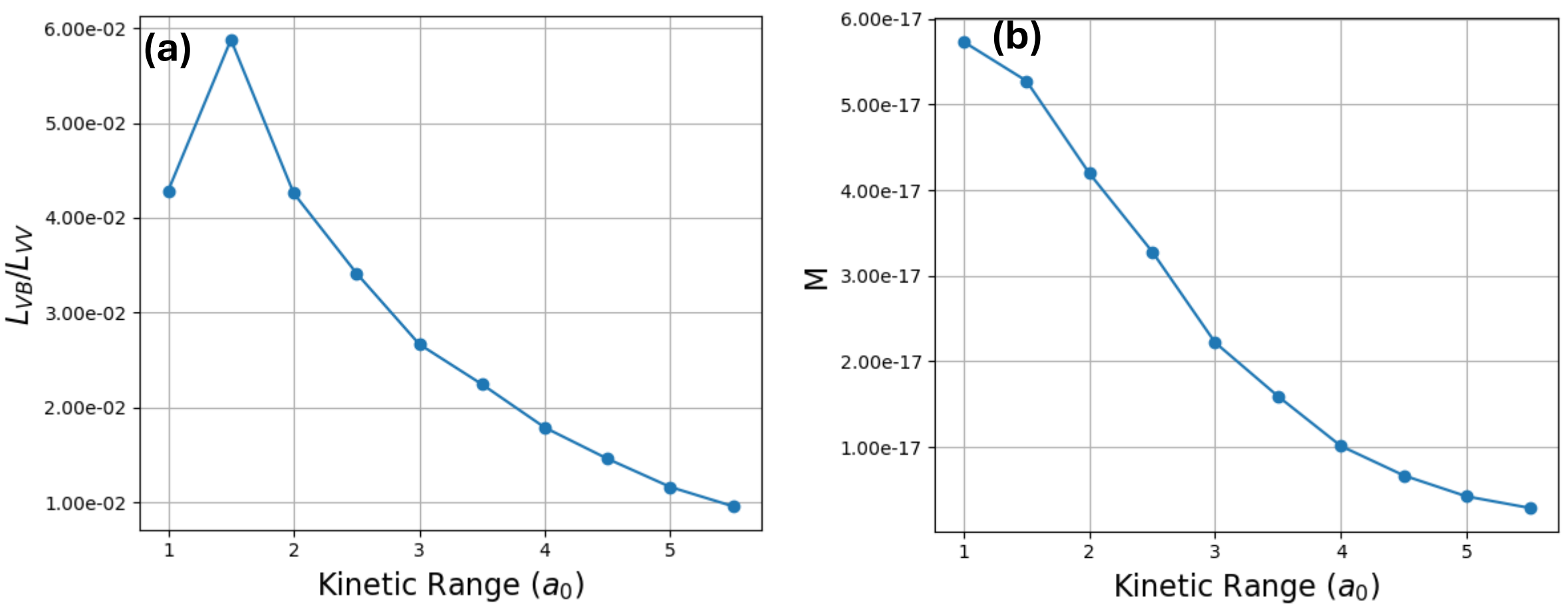

Figure S2**:** Convergence of kinetic range with respect to $L_{VB}/L_{VV}$ in (a) and Cluster Mobility in (b) for the V-Zr-Fe cluster using 1NN thermodynamic range with pairwise triplet binding energies and the KRA method for estimating jump frequencies.

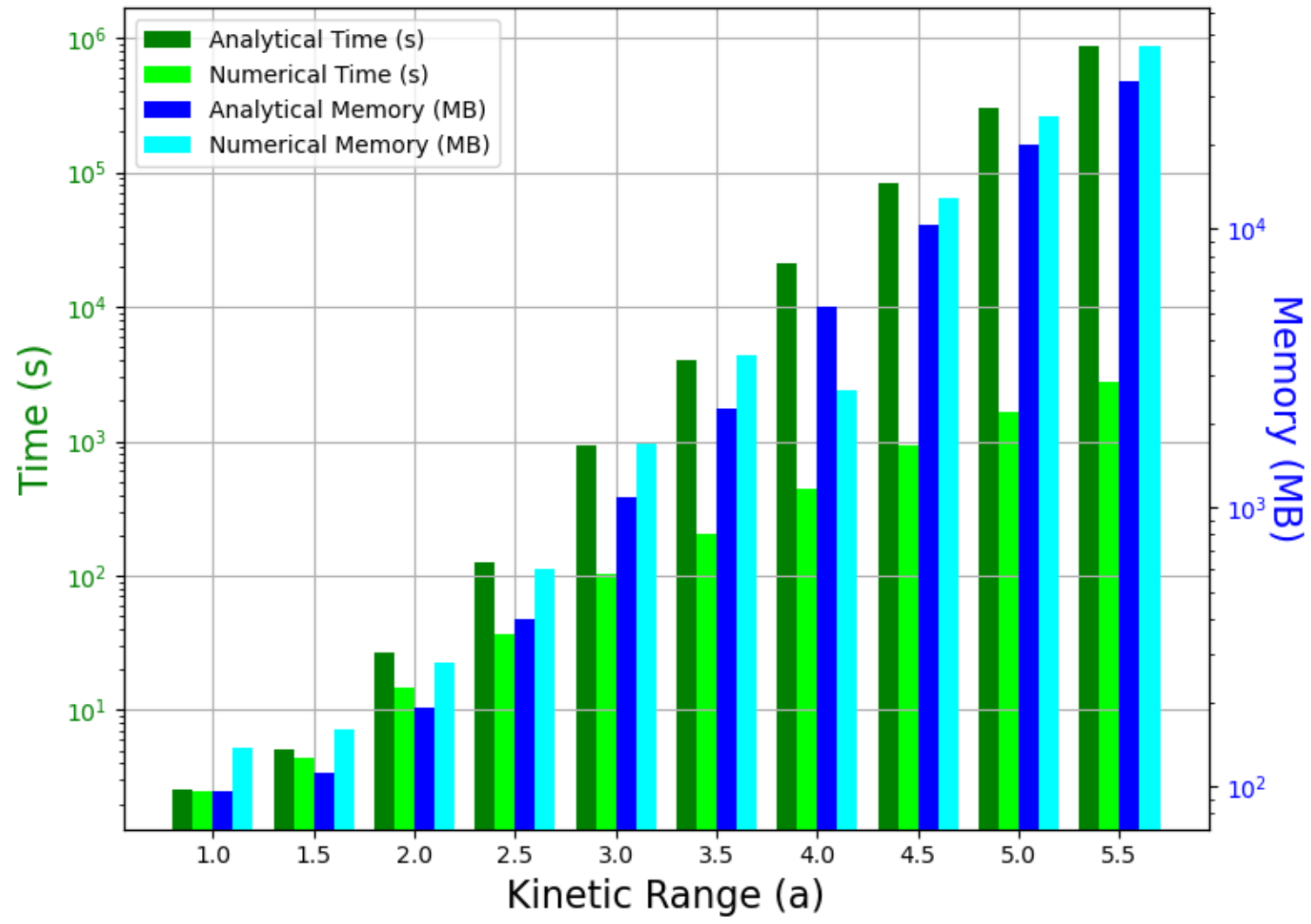

Figure S3: KineCluE computational time and memory associated with V-Fe-Zr triplet calculations using a kinetic range from $1.0a_0$ to $5.5a_0$ and a fixed thermodynamic range of 1NN.

### 3. Sensitivity analysis of triplet V-Zr-Fe and V-Zr-Co clusters

The sensitivity analysis feature in KineCluE [4] was employed to obtain well-converged transport coefficients of the V-Zr-Fe and V-Zr-Co clusters with respect to DFT migration energies. This is accomplished by identifying which migration energies within the V-Zr-Fe clusters most significantly influence the relevant transport properties of the system ($L_{BV}$, $L_{VV}$, and $L_{BB}$). First, the triplet binding energies with 1NN Zr-C and 1NN Zr-V pairs are calculated using DFT since these configurations are predicted to have the highest binding energies from the sum of the constituent pairwise interactions and therefore are likely to have the largest impact on the cluster transport coefficients. The rest are estimated using additive pairwise binding energies between V-Zr, Zr-*C*, and V-C since their predicted binding energies are significantly lesser than the aforementioned. This set of triplet binding energies is used in KineCluE to form an estimate of migration energies using the kinetically resolved approximation (KRA). KineCluE then uses this set of migration energies to calculate the transport coefficients of the cluster and to find the most important migration energies using the sensitivity analysis feature. These migration energies are then accurately calculated using DFT as shown in Figure S4 and Figure S5. The triplet migration energies are notated in the subscript as two triplet pairs with the NN distances between V-Zr, V-C, and Zr-C, before and after the jump, respectively (e.g. $\omega_{111,111}^{Ex-Zr}$ for vacancy swap with Zr atom with all 1NN pair interactions in the triplet cluster). Additionally, the exchanging atom type is notated as a superscript if not a matrix atom. Within the first iteration of this sensitivity analysis (e.g. within two jump frequencies), the transport coefficients converged quite well as shown in Figure S14 for the V-Zr-Fe cluster. The most important jump frequencies in the triplet clusters seem to be heavily influenced by the strong *B*-*C* binding energy. Ultimately, there are 2 configurations and 5 jump frequencies for the V-*B* systems (e.g. the five jump frequency model) which are all calculated using DFT in Table S1. For the V-*B*-*C* systems, this kinetic and thermodynamic range results in 14 configurations and 53 jump frequencies which shows the need for the sensitivity study in KineCluE. For the triplet clusters, the 4 most important triplet configurations (highest predicted binding energy by constituent pairwise interactions) are calculated with DFT.

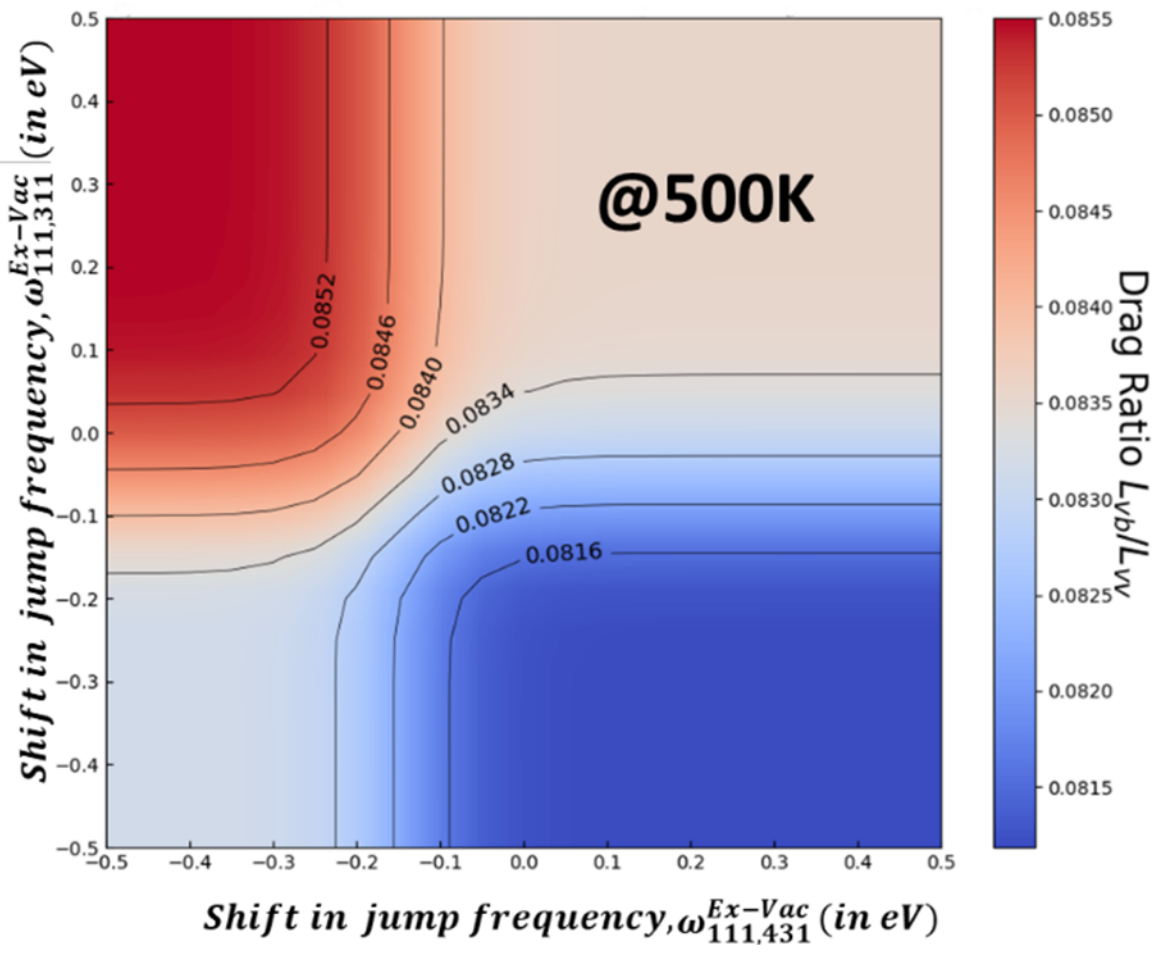

Figure S4: Sensitivity study performed on jump frequencies within the V-Zr-Fe cluster after inputting triplet DFT binding energies

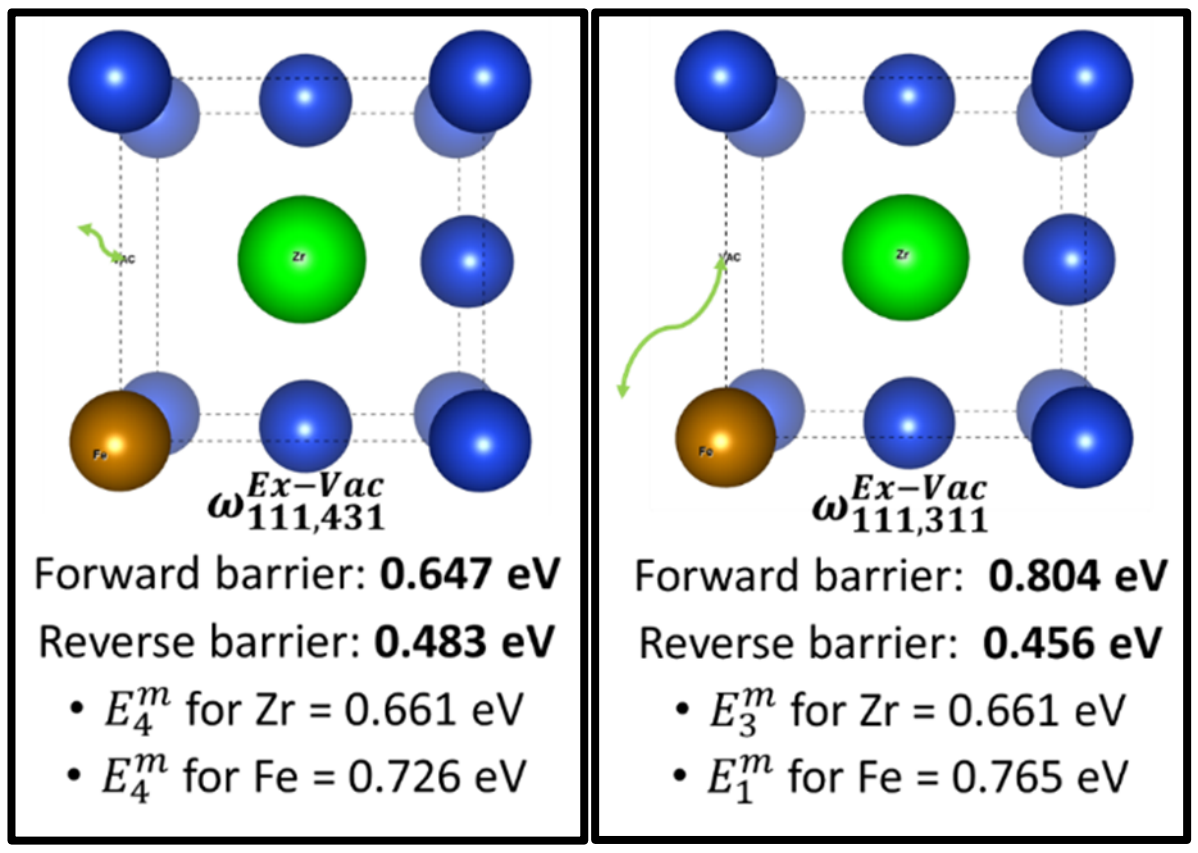

Figure S5: DFT-calculated migration energies for V-Zr-Fe cluster compared to jump frequencies for the binary systems

## 4. Electron Density Maps of clusters in FCC Cu

The valence electron density maps are shown in Figure S6 for various solute and defect pairs within a Cu FCC matrix, computed using DFT. These maps represent a two-dimensional slice of the (111) plane, providing insights into the local bonding environment around solute-solute and solute-defect pairs in the copper matrix. The configurations analyzed include V-Zr, V-Fe, Fe-Fe, Co-Co, and Zr-Zr pairs. The Fe-Fe and Co-Co attractive pairs show strong directional bonding with a region of elevated electron density between the pairs, while the Zr-Zr repulsive pairs show a region of lower electronic density between the Zr-Zr pairs. In addition, the interatomic distances are consistent with the size mismatch of the solutes in the copper matrix. For example, when Zr is present as an isolated solute in the Cu matrix, the Zr-Cu interatomic distance is 2.67 Å, which is noticeably longer than the 2.57 Å Cu-Cu interatomic distance in pure Cu, reflecting that Zr is an oversized solute in Cu with weaker solute-matrix interactions. In contrast, within the (111) triangle configuration of the V-Zr-C clusters, the Zr-Co interatomic distance reduces to 2.32 Å, which is even shorter than the 2.40 Å observed for Zr-Fe, indicating stronger directional bonding between Zr-Co consistent with the smaller atomic size of Co compared to Fe.

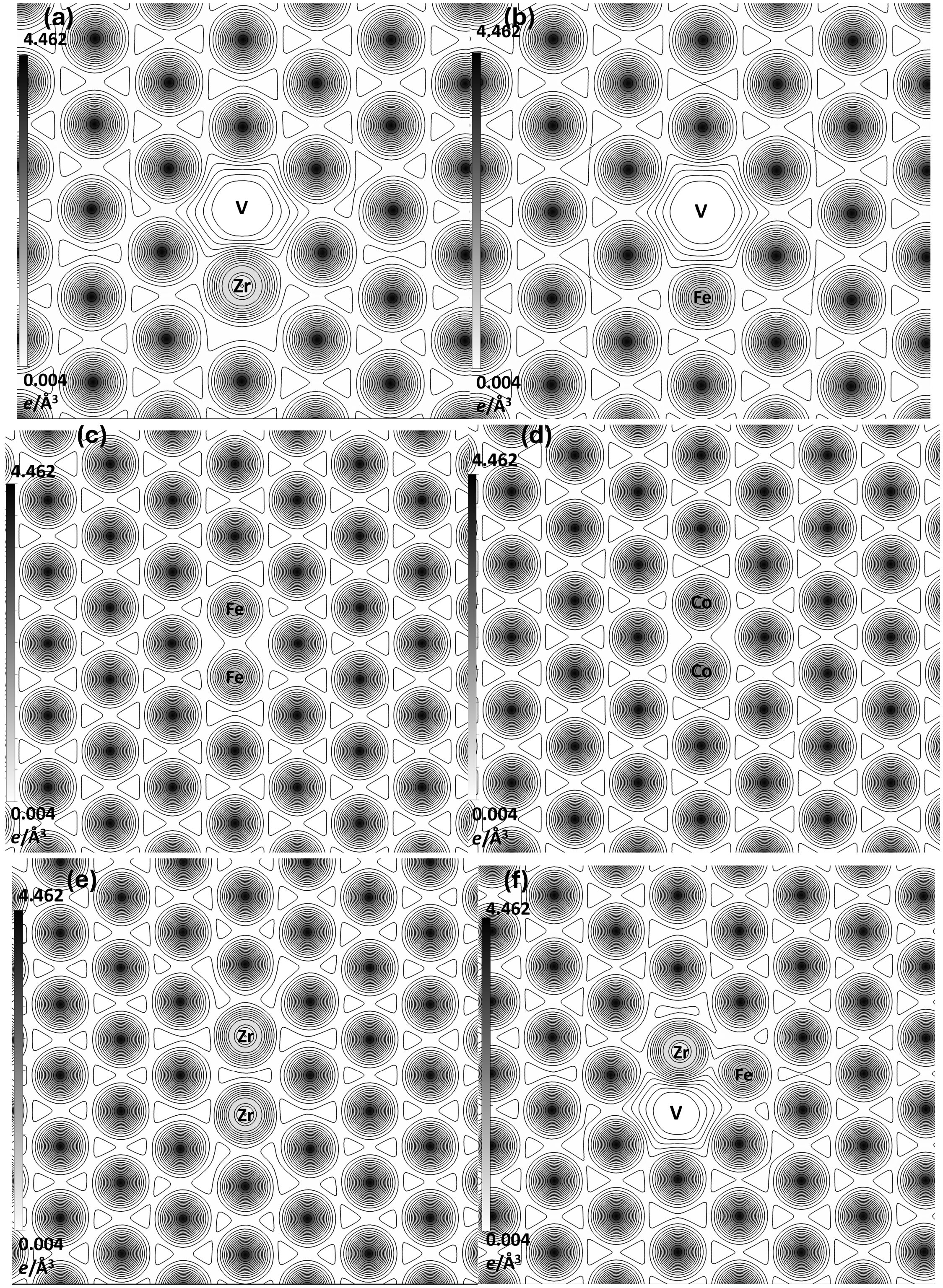

Figure S6: Valence electron density maps of (a) V-Zr, (b)V-Fe, (c) Fe-Fe, (d) Co-Co, (e) Zr-Zr , and (f) V-Zr-Fe clusters in a Cu FCC matrix computed using DFT. A slice of the (111) plane is shown. Contour lines with logarithmic intervals in units of e/Å$^3$ are shown. Matrix Cu atoms are not labeled.

## 5. Convergence testing for DFT calculations

Vacancy formation energy was calculated for various k-point meshes and smear degauss values to test the convergence of the DFT calculations, see Figure S7. The chosen k-point mesh (4x4x4) and degauss value (2 mRy or 0.027 eV) are within 10 meV of the fully-converged 8x8x8 k-point mesh calculations. The energies calculated with Methfessel-Paxton smearing function are relatively independent of degauss value when converged with respect to k-point density.

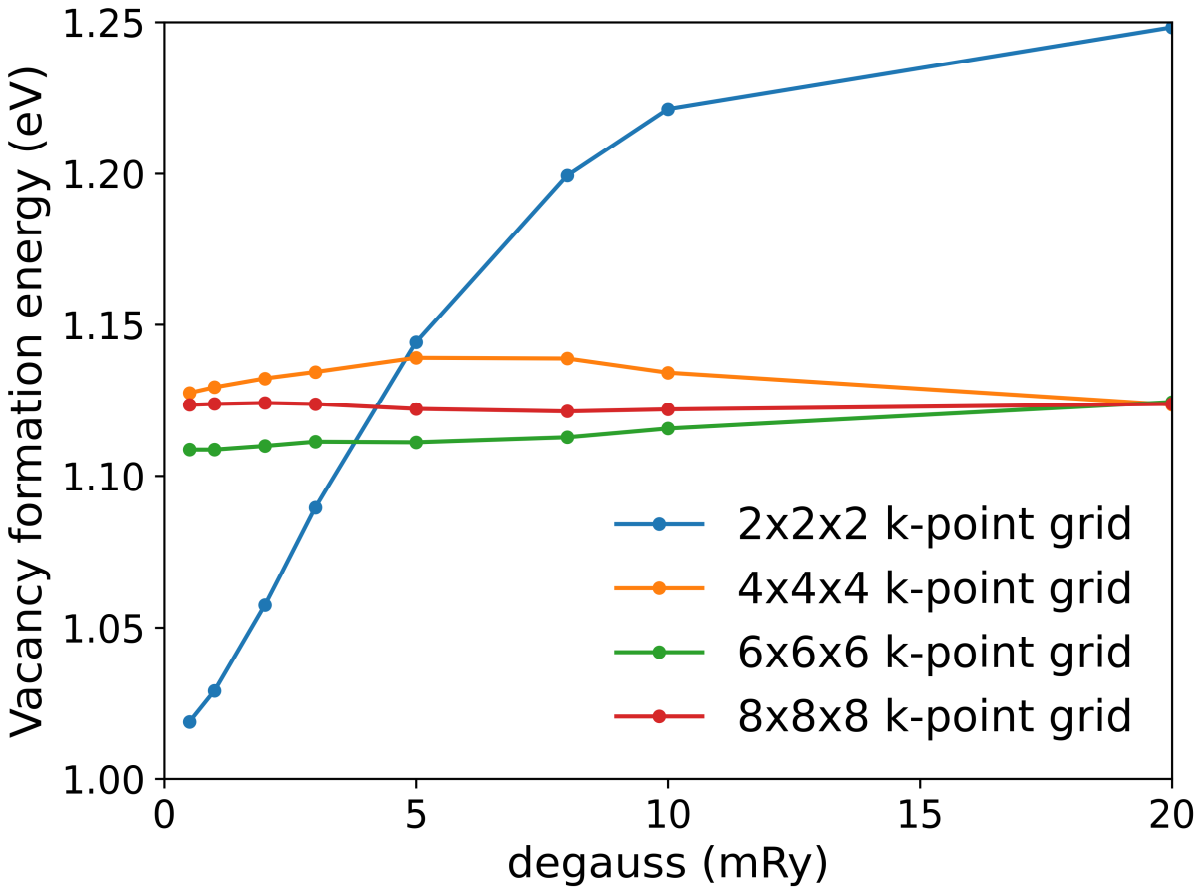

Figure S7: Convergence of vacancy formation energy with respect to k-point grid and smearing degauss using a 3x3x3 supercell containing 107 or 108 Cu atoms